\documentstyle[12pt]{article}

\begingroup\makeatletter\ifx\SetFigFont\undefined
\def\x#1#2#3#4#5#6#7\relax{\def\x{#1#2#3#4#5#6}}%
\expandafter\x\fmtname xxxxxx\relax \def\y{splain}%
\ifx\x\y   
\gdef\SetFigFont#1#2#3{%
  \ifnum #1<17\tiny\else \ifnum #1<20\small\else
  \ifnum #1<24\normalsize\else \ifnum #1<29\large\else
  \ifnum #1<34\Large\else \ifnum #1<41\LARGE\else
     \huge\fi\fi\fi\fi\fi\fi
  \csname #3\endcsname}%
\else
\gdef\SetFigFont#1#2#3{\begingroup
  \count@#1\relax \ifnum 25<\count@\count@25\fi
  \def\x{\endgroup\@setsize\SetFigFont{#2pt}}%
  \expandafter\x
    \csname \romannumeral\the\count@ pt\expandafter\endcsname
    \csname @\romannumeral\the\count@ pt\endcsname
  \csname #3\endcsname}%
\fi
\fi\endgroup

\advance \topmargin by -\headheight
\advance \topmargin by -\headsep

\evensidemargin \oddsidemargin
\begin{document}

\newpage
\pagestyle{empty}

\begin{flushright}
CERN-TH/99-70\\
gr-qc/9903080 \\
$~$ \\
March 1999
\end{flushright}

\begin{centering}

\baselineskip 21pt plus 0.2pt minus 0.2pt

\bigskip
{\Large {\bf Gravity-wave interferometers as probes of a \\
low-energy effective quantum gravity} }
 
\bigskip

\baselineskip 12pt plus 0.2pt minus 0.2pt

\bigskip
\bigskip
\bigskip
\bigskip
 
{\bf Giovanni AMELINO-CAMELIA}\footnote{{\it Marie Curie
Fellow} of the European Union }\\
\bigskip
Theory Division, CERN, CH-1211, Geneva, Switzerland

\end{centering}
\vspace{2cm}
     
\centerline{\bf ABSTRACT }
\medskip

\begin{quote}

The interferometry-based experimental tests of 
quantum properties of space-time 
which the author sketched out
in a recent short Letter [Nature 398 (1999) 216]
are here discussed in
self-contained fashion. 
Besides providing detailed derivations of the results
already announced in the previous Letter, 
some new results are also derived; in particular,
the analysis is extended to a larger class of scenarios 
for space-time fuzziness and an absolute bound on the measurability
of the amplitude of a gravity wave is obtained.
It is argued that these studies could be
helpful for the search of a theory describing a first stage 
of partial unification of Gravity and Quantum Mechanics.

\end{quote}
\baselineskip 18pt

\vfill

\newpage
\pagenumbering{arabic}
\setcounter{page}{1}
\pagestyle{plain}
\baselineskip 12pt plus 0.2pt minus 0.2pt

\section{INTRODUCTION AND SUMMARY}

Perhaps the most fascinating questions confronting
contemporary physics concern the search of the appropriate
framework for the unified description of Gravity and Quantum Mechanics.
This search for ``Quantum Gravity''
is proving very difficult~\cite{nodata}, especially as a result
of the scarce 
experimental information available on the interplay 
between Gravity and Quantum Mechanics. 
However, in recent years there has been a small (but nevertheless
encouraging) number of new proposals~\cite{elmn,venegwi,grbgac,ahluexp}
of experiments probing the nature of the interplay between 
Gravity and Quantum Mechanics. At the same time the ``COW-type'' 
experiments,
initiated with the celebrated experiment
by Colella, Overhauser and Werner~\cite{cow},
have reached levels of sophistication~\cite{chu} such that
even gravitationally induced quantum phases 
due to local tides can be detected.
In light of these developments there 
is now growing (although still understandably
cautious) hope for data-driven 
insight into the structure of Quantum Gravity.

The primary objective of the
present Article is the one of providing
a careful discussion of 
the most recent addition to the (still far from numerous)
family of Quantum Gravity experiments,
which this author proposed in the short 
Letter in Ref.~\cite{gacgwi}. 
This most recent proposal probes in a rather direct way the properties 
of space-time, which is of course the most fundamental element of a 
Quantum Gravity, by exploiting the remarkable accuracy achievable with 
advanced modern interferometers, such as the ones
used for searches of gravity waves.

While perhaps (especially in light of the gloom overall
status of ``Quantum Gravity phenomenology'')
already sufficient
interest in the experiment proposed in Ref.~\cite{gacgwi}
could come from a pragmatic
phenomenological viewpoint, in this Article I shall also
relate the class of observations accessible to modern interferometers
to a physical picture of the (necessarily small) way in which
Quantum Gravity might affect phenomena probing space-time
at distances significantly larger than the Planck 
length $L_{planck} \sim 10^{-35}m$
(but significantly shorter than distance scales probed 
in ordinary particle-physics or gravity experiments).
This physical picture is motivated by the 
huge gap between the minute Planck length
and the distance scales probed 
in present-day particle-physics or gravitational experiments. 
The size of this gap provides motivation for exploring the
possibility that on the way to
Planck-length physics a few intermediate steps of partial unification
of Gravity and Quantum Mechanics might be required before reaching
full unification. Of course, as long as we are lacking direct
experimental evidence to the contrary, it is also reasonable
to work (as many distinguished colleagues do)
on the hypothesis that Gravity and Quantum Mechanics should
merge directly into a fully developed Quantum Gravity,
but in the present Article (as in the previous 
papers \cite{gacmpla,gacxt,gacgrf98}) I shall be concerned
with the investigation of the properties that 
one could demand of a theory
suitable for a first stage of partial
unification of Gravity and Quantum Mechanics.
In particular, I shall review the arguments presented in
Refs.~\cite{gacxt,gacgrf98} suggesting that 
the most significant implications
of Quantum Gravity for low-energy (large-distance)
physics might be associated to
the structure of the non-trivial ``Quantum Gravity vacuum''.
A satisfactory picture of this Quantum Gravity vacuum
is not available at present, and therefore we must generically characterize 
it as the appropriate new concept that in Quantum Gravity
takes the place of the ordinary concept of ``empty space'';
however, it is plausible that some of 
the arguments by Wheeler, Hawking
and others (see, {\it e.g.}, Refs.~\cite{wheely,hawk} and
references therein),
which have attempted to develop an intuitive description
of the Quantum Gravity vacuum, might have captured 
at least some of its actual properties.

Other possible elements for the search of a theory
suitable for a first stage of partial
unification of Gravity and Quantum Mechanics
come from studies suggesting that this unification might
require a novel relationship between ``measuring apparatus''
and ``system''. My intuition on the nature
of this new relationship is mostly based on work
by Bergmann and Smith~\cite{bergstac}
and the observations I reported in Refs.~\cite{gacmpla,gacgrf98},
which took as starting point an analysis
by Salecker and Wigner~\cite{wign}.

The intuition emerging from these considerations on
a novel relationship between measuring apparatus
and system and by a Wheeler-Hawking picture
of the Quantum Gravity vacuum
are not sufficient for the full development of a new
formalism describing the first stage of partial
unification of Gravity and Quantum Mechanics,
but they provide encouragement for the search
of a formalism based on a mechanics not exactly
of the type of ordinary Quantum Mechanics. Moreover,
one can use this emerging intuition for rough estimates
of certain candidate Quantum-Gravity effects.
The estimates most relevant for the present Article 
are the ones concerning the space-time ``fuzziness''
which modern interferometers could investigate 
following Ref.~\cite{gacgwi}.

A prediction of nearly
all approaches to the unification
of Gravity and Quantum Mechanics is that 
at very short distances the sharp
classical concept of space-time should give way 
to a somewhat ``fuzzy'' (or ``foamy'') 
picture (see, {\it e.g.}, Refs.~\cite{wheely,hawk,arsarea,emn}),
but it is usually very hard to characterize this
fuzziness in physical operative terms.
In Section~2 
I provide an operative definition of fuzzy distance
that has completely general applicability.
My operative definition of fuzzy distance involves the use of 
interferometers, and the remarkable
recent progress in the accuracy of these devices
provides motivation for an analysis aimed at investigating the possible
observable implications of Quantum Gravity for
modern interferometers.
In Section~3 I provide estimates for the
quantum fluctuations that could affect
distances if the above-mentioned intuition
on the first stage of partial
unification of Gravity and Quantum Mechanics is correct.
I shall proceed with the attitude
of searching for plausible (but admittedly ``optimistic'')
estimates of the relevant Quantum Gravity effects, and, although
quantitative estimates will be derived, the true emphasis is on
the qualitative aspects of the phenomena,
since this type of information could be helpful
to colleagues on the experimental side in establishing how to look
for these phenomena.
Some of the estimates I provide are motivated by
studies of the measurability
of distances in Quantum Gravity.
A second group of  estimates is based on
elementary toy models of the stochastic processes that might
characterize space-time fuzziness.
The third and final group of estimates is motivated
by arguments of ``consistency''
(in the sense discussed later) with recent 
proposals~\cite{grbgac,aemn1,gampul}
of Quantum-Gravity induced deformation of the dispersion
relation that characterizes the propagation of massless particles.
All of these arguments indicate that a priority for interferometry-based
tests of space-time fuzziness must be high sensitivity at
low frequencies, and I hope this will
be taken into account in planning
future gravity-wave interferometers.

In Section~4 I shall observe (extending the related observations
reported in Ref.~\cite{gacgwi}) that the remarkable sensitivity
achieved by modern 
interferometers, especially the ones used to search 
for gravity waves~\cite{saulson,ligoprototype,ligo,virgo,lisa},
allows to set highly significant bounds on some of the
fuzziness scenarios discussed in Section~2.
Perhaps the most intuitive way to characterize the obtained bounds
is given by the fact that we are now in a position to rule out
a picture of fuzzy space-time such that minute Planck-length
($10^{-35} m$)
fluctuations would affect distances at a rate of one per
each Planck time $10^{-44} s$.
In Section~5 I derive a novel absolute bound on the measurability
of the amplitude of a gravity wave.
This measurability bound is obtained by combining
a well-known ``standard quantum limit,'' which depends
on the mass of the mirrors used by the gravity-wave
interferometers,
and a limitation on the mass of the mirrors that
is imposed by gravitational effects.
I find that this measurability bound is too weak to be tested 
with available or planned gravity-wave interferometers.
Its significance mostly resides in the fact that it illustrates
even more clearly than previous measurability analyses
the fact that the
unification of Gravity and Quantum Mechanics
requires a new relationship between measuring apparatus
and system. 
In Section~6 I discuss the aspects 
of certain existing Quantum Gravity approaches
which are in one or another way 
related to
the type of fuzzy space-times considered in Section~2.
In Section~7 I discuss how the class of experiments
proposed in Ref.~\cite{gacgwi} (and here analyzed in detail)
complements other proposals of Quantum Gravity experiments.
I also outline the general features that an experiment must
have in order to uncover aspects of the 
interplay between Gravity and Quantum Mechanics.
In Section~8 I use the results discussed in Sections 2,3,4,5,6
to better define the idea of a theory appropriate
for the description of a first stage of partial
unification of Gravity and Quantum Mechanics.
Closing remarks, also on the outlook for
Quantum-Gravity phenomenology,
are offered in Section~9.

\section{OPERATIVE DEFINITION OF FUZZY DISTANCE}

While nearly all approaches to the unification
of Gravity and Quantum Mechanics appear to lead
to a somewhat fuzzy
picture of space-time, 
within the various formalisms it is often difficult to
characterize physically this fuzziness.
Rather than starting from formalism, I shall advocate
an operative definition of fuzzy space-time.\footnote{Once we 
have a physical definition of fuzzy space-time the analysis 
of the various Quantum Gravity formalisms could be aimed at 
providing predictions for this fuzziness. Of course, 
in order for the formalisms to provide such physical predictions
it is necessary to equip them with at least some elements of
a ``measurement theory''.}
More precisely for the time being I shall just consider 
the concept of fuzzy distance.
I shall be guided by the expectation that
at very short distances the sharp
classical concept of distance should give way 
to a somewhat fuzzy distance. Since interferometers
are ideally suited to monitor the distance between
test masses, I choose as operative definition of 
Quantum-Gravity induced fuzziness 
one which is expressed in terms
of Quantum-Gravity induced noise in the 
read-out of interferometers.

In order to articulate this proposal it will prove useful
to briefly review some aspects of the physics of 
Michelson interferometers.
These are schematically composed~\cite{saulson}
of a (laser) light source, a beam splitter 
and two fully-reflecting mirrors placed
at a distance
$L$ from the beam splitter in orthogonal directions.
The light beam is decomposed by the beam splitter 
into a transmitted beam directed toward one of the mirrors
and a reflected beam directed toward the other mirror;
the beams are then reflected by the mirrors 
back toward the beam splitter,
where~\cite{saulson} they are 
superposed\footnote{Although all modern interferometers
rely on the technique of folded interferometer's arms
(the light beam bounces several times between the
beam splitter and the mirrors before superposition),
I shall just discuss 
the simpler ``no-folding'' conceptual setup.
The readers familiar with the subject can easily realize
that the observations here reported also apply
to more realistic setups, although in some steps of the derivations
the length $L$ would have to be understood as the optical length
(given by the actual length of the arms times the number
of foldings).}.
The resulting interference pattern 
is extremely sensitive to changes in the positions of the mirrors
relative to the beam splitter.
The achievable
sensitivity is so high that planned interferometers~\cite{ligo,virgo}
with arm lengths $L$ of $3$ or $4$ $Km$
expect to detect gravity waves of amplitude $h$ as 
low as $3 \cdot 10^{-22}$ at frequencies of about $100 Hz$.
This roughly means that these modern gravity-wave interferometers
should monitor the (relative) positions of their test masses
(the beam splitter and the mirrors) with an accuracy~\cite{ligoprototype}
of order $10^{-18} m$ and better.

In achieving this remarkable accuracy experimentalists must
deal with classical-physics displacement noise sources ({\it e.g.},
thermal and seismic effects induce fluctuations in the relative
positions of the test masses) and displacement noise sources
associated to effects of ordinary Quantum Mechanics
(as I shall mention again later the combined minimization
of {\it photon shot noise} and {\it radiation pressure noise}
leads to an irreducible noise source which has its root in
ordinary Quantum Mechanics).
The operative definition of fuzzy distance which I advocate
characterizes the corresponding Quantum Gravity effects
as an additional source of displacement noise.
A theory in which the concept of distance is 
fundamentally fuzzy in this 
operative sense would be such that even in the idealized
limit in which all classical-physics and ordinary Quantum-Mechanics
noise sources are completely eliminated the read-out of an
interferometer would still be noisy as a result of Quantum 
Gravity effects.

Adopting this operative definition of fuzzy distance,
interferometers are of course the natural tools for 
experimental tests of proposed space-time fuzziness scenarios.
However, even the remarkable sensitivity
estimate of order $10^{-18} m$ given above is quite far
from the Planck length $\sim 10^{-35}m$, and it might appear
safe to assume that any scenario for space-time fuzziness
would not observably affect the operation of even the most 
sophisticated modern interferometers.
In spite of the intuition emerging from this preliminary considerations,
in the next Sections~3 and 4 I shall show that some plausible (albeit
somewhat speculative) fuzziness scenarios can be tested
in a rather significant way by modern interferometers. 
The key observation 
is based on the fact that the physics of an interferometer
involves other length scales besides the $10^{-18} m$ length scale
discussed above, and the combinations of length scales
which characterize on the one hand the noise levels achievable by
modern interferometers and on the other hand the Quantum-Gravity
induced noise levels turn out to be comparable.
In particular, a proper description of noise levels in an interferometer
must provide the displacement sensitivity as a function
of frequencies $f$ (notice the additional length scale $c f^{-1}$ obtained
combining $f$ with the speed-of-light constant $c\sim 3 \cdot 10^8 m/s$),
and similarly the ``amount of fuzziness'' 
predicted by certain 
space-time fuzziness scenarios turns out to be $f$-dependent.
Within certain ranges of values of $f$ one finds that the
experimental limits are actually significant with respect to
the theoretical predictions.
Before providing this phenomenological analysis I shall use the next
Section to discuss estimates of the type of noise levels
that could be expected within certain space-time fuzziness scenarios.

\section{SOME CANDIDATE FUZZY SPACE-TIMES}

\subsection{Minimum-length noise}

In many Quantum Gravity approaches there appears to be
a length scale $L_{min}$, often identified with 
the string length ($L_{string} \sim 10^{-34}m$)
or the Planck length, which sets an absolute bound
on the measurability of distances (a minimum uncertainty):
\begin{eqnarray}
\delta D \ge L_{min}
~. \label{minlen}
\end{eqnarray}
This property emerges in approaches based on canonical
quantization of Einstein's gravity when analyzing
certain gedanken experiments
(see, {\it e.g.}, Ref.~\cite{padma,garay} and references therein).
In Critical Superstring Theories, theories whose mechanics is 
still governed
by the laws of ordinary Quantum Mechanics but with one-dimensional
(rather than point-like) fundamental objects,
a relation of type (\ref{minlen}) follows from
the stringy modification
of Heisenberg's uncertainty principle~\cite{venekonmen}
\begin{eqnarray}
 \delta x \, \delta p \!\!& \ge &\!\! 1
+ {L_{string}^2} \, \delta p^2
~. \label{veneup}
\end{eqnarray}
In fact, whereas Heisenberg's uncertainty principle
allows $\delta x = 0$ (for $\delta p \rightarrow \infty$),
for all choices of  $\delta p$ 
the uncertainty relation (\ref{veneup})
gives $\delta x \ge L_{string}$.
The relation (\ref{veneup}) is suggested by certain analyses
of string scattering~\cite{venekonmen}, but it
might have to be modified when taking into account
the non-perturbative solitonic structures of Superstring Theory
known as Dirichlet branes~\cite{dbrane}.
In particular, evidence has been found~\cite{dbrscatt}
in support of the possibility that ``Dirichlet
particles" (Dirichlet 0~branes)
could probe the structure of space-time down
to scales shorter than the string length.
In any case, all evidence available on Critical Superstring Theory
is consistent with a relation of type (\ref{minlen}),
although it is probably safe to say that
some more work is still needed to firmly establish 
the string-theory value of $L_{min}$.

Having clarified that a relation of type (\ref{minlen})
is a rather common prediction of theoretical work
on Quantum Gravity, let us then consider how such a relation
could affect the noise levels of an interferometer,
{\it i.e.} let us consider the type of fuzziness
(in the sense of the operative definition I advocated)
which could be encoded in relation (\ref{minlen}).

First let us observe that relation (\ref{minlen})
does not necessarily encode any fuzziness; for example,
relation (\ref{minlen}) could simply emerge from
a theory based on a lattice of points with spacing $L_{min}$
and equipped with a measurement theory consistent
with (\ref{minlen}).
The concept of distance in such a theory
would not necessarily be affected by the
type of stochastic processes that lead to noise in
an interferometer.

However, it is also possible for relation (\ref{minlen})
to encode the net effect of some underlying physical processes
of the type one would qualify as quantum space-time fluctuations.
These fluctuations, following work initiated by Wheeler and Hawking,
are often visualized
as involving geometry and topology fluctuations~\cite{wheely},
virtual black holes~\cite{hawk},
and other novel phenomena.
A very intuitive description of the way in which
the dynamics of matter distributions would be affected
by this type of fuzziness of space-time
is obtained by noticing certain similarities~\cite{garaythermal}
between a thermal environment and the environment
of quantum space-time fluctuations consistent with (\ref{minlen}).
This (however preliminary) network of intuitions suggests that
(\ref{minlen}) could be the result of fuzziness for distances $D$
of the type associated to stochastic fluctuations with
root-mean-square deviation $\sigma_D$ given by
\begin{equation}
\sigma_D \sim L_{min} \, .
\label{no1}
\end{equation}
The associated
displacement amplitude spectral density $S_{min}(f)$
should roughly have a $1/\sqrt{f}$ behaviour
\begin{equation}
S_{min}(f) \sim {L_{min} \over \sqrt{f}} \, .
\label{no1spectrum}
\end{equation}
This can be justified using the observation
that for a frequency-band limited from below only
by the time of observation $T_{obs}$
the relation
between $\sigma$
and $S(f)$ is given by~\cite{rwold}
\begin{eqnarray}
\sigma^2 = \int_{1/T_{obs}}^{f_{max}}
[S(f)]^2 \,  df ~.
\label{gacspectrule}
\end{eqnarray}
Substituting the $S_{min}(f)$ of Eq.~(\ref{no1spectrum})
for the $S(f)$ of Eq.~(\ref{gacspectrule})
one obtains a $\sigma$ that approximates the $\sigma_D$
of Eq.~(\ref{no1}) up to small (logarithmic) $T_{obs}$-dependent
corrections.
A more detailed description of the
displacement amplitude spectral density associated
to Eq.~(\ref{no1}) can be found in Refs.~\cite{jare1,jare2}.
For the objectives of the present article the rough
estimate (\ref{no1spectrum}) is sufficient since,
if indeed $L_{min} \sim L_{planck}$, from (\ref{no1spectrum}) 
one obtains $S_{min}(f) \sim 10^{-35} m / \sqrt{f}$,
which is still very far from the sensitivity of even the most 
advanced modern
interferometers, and therefore we should not be concerned with
corrections to Eq.~(\ref{no1spectrum}).

\subsection{Random-walk noise motivated by the analysis of
a Salecker-Wigner gedanken experiment}

The above argument relating the measurability bound (\ref{minlen})
to fuzziness of type (\ref{no1}) can be used in general to relate 
any bound on the measurability of distances to an estimate
of the possible stochastic quantum fluctuations affecting
the operative definition of distances.
In this Subsection 3.2 I shall consider a measurability
bound that emerges when taking into account the quantum
properties of devices.
It is well understood
(see, {\it e.g.},
Refs.~\cite{gacmpla,gacgrf98,bergstac,diosi,ng,dharam94grf}) 
that the combination of 
the gravitational properties 
and 
the quantum
properties 
of devices can have an important
role in the analysis
of the operative definition of gravitational observables.
Since the analyses~\cite{padma,garay,venekonmen,dbrscatt}
that led to the proposal of Eq.~(\ref{no1})
only treated the devices in a completely idealized manner
(assuming that one could ignore any contribution to
the uncertainty in the measurement of $D$ due to the
gravitational and quantum properties of devices),
it is not surprising that analyses that took into account the 
gravitational and quantum properties of devices found more 
significant limitations to the measurability of distances.

Actually, by ignoring the way in which the gravitational properties 
and the quantum properties of devices combine in measurements
of geometry-related physical properties of a system
one misses some of the fundamental elements
of novelty we should expect for the interplay of Gravity
and Quantum Mechanics; in fact, one would be missing an
element of novelty which is deeply associated to the Equivalence
Principle.
In measurements of physical properties
which are not geometry-related one can safely
resort to an idealized description of devices.
For example, in
the famous Bohr-Rosenfeld analysis~\cite{rose}
of the measurability of the electromagnetic field
it was shown that the accuracy allowed by
the formalism of ordinary Quantum Mechanics could only be achieved
using idealized test particles with vanishing ratio between
electric charge and inertial mass.
Attempts to generalize the Bohr-Rosenfeld analysis
to the study of gravitational fields
(see, {\it e.g.}, Ref.~\cite{bergstac})
are of course confronted with the fact that
the ratio between gravitational ``charge'' (mass) and inertial mass
is fixed by the Equivalence Principle.
While ideal devices with vanishing ratio between  
electric charge and inertial mass can
be considered at least in principle,
devices with vanishing ratio between  
gravitational mass and inertial mass 
are not admissible in any (however formal) limit
of the laws of gravitation.
This observation provides one of the strongest elements
in support of the idea~\cite{gacgrf98}
that the mechanics on which Quantum
Gravity is based must not be exactly
the one of ordinary Quantum Mechanics, since it should
accommodate a somewhat different relationship between ``system''
and ``measuring apparatus''. [In particular, the new mechanics
should not rely on the idealized ``measuring apparatus''
which plays such a central role in the mechanics laws of
ordinary Quantum Mechanics, see, {\it e.g.}, 
the ``Copenhagen interpretation''.]

In trying to develop some intuition for the type
of fuzziness that could affect the concept of distance
in Quantum Gravity, it might be useful to consider the way 
in which the interplay between the
gravitational and the quantum properties of devices affects
the measurability of distances.
In Refs.~\cite{gacmpla,gacgrf98} I have argued that
a natural starting point for this type of analysis
is provided by the procedure for the measurement of distances
which was discussed in 
influential work by Salecker and Wigner~\cite{wign}.
These authors ``measured'' (in the ``{\it gedanken}'' sense)
the distance $D$ between two bodies 
by exchanging a light signal between them.
The measurement procedure requires
by {\it attaching}\footnote{Of course,
for consistency with causality,
in such contexts one assumes devices to be ``attached non-rigidly,''
and, in particular, the relative position
and velocity of their centers of mass continue to satisfy the
standard uncertainty relations of Quantum Mechanics.} 
a light-gun ({\it i.e.} a device 
capable of sending
a light signal when triggered), a detector
and a clock to
one of the two bodies 
and {\it attaching} a mirror to the other body. 
By measuring the time $T_{obs}$ (time of observation)
needed by the light signal
for a two-way journey between the bodies one 
also
obtains a 
measurement of  
the distance $D$.
For example, in Minkowski space 
and neglecting quantum effects 
one simply finds that $D = c {T_{obs} / 2}$.
Within this setup it is easy to realize that the
interplay between the
gravitational and the quantum properties of devices 
leads to an irreducible contribution to the uncertainty $\delta D$.
In order to see this it is sufficient to consider the
contribution to $\delta D$ coming from 
the uncertainties that affect the motion
of the center of mass of the system
composed by the light-gun, the detector and the clock.
Denoting with $x^*$ and $v^*$
the position and the velocity of the center of mass
of this composite device
relative to the position of the body to which it is {\it attached},
and assuming that the experimentalists prepare this device
in a state characterised by
uncertainties $\delta x^*$ and $\delta v^*$,
one easily finds~\cite{wign,gacgrf98}
\begin{eqnarray}
\delta D \geq 
\delta x^* + T_{obs} \delta v^* 
\geq 
\delta x^* 
+ \left( {1 \over  M_b} + {1 \over  M_d} \right)
{ \hbar T_{obs} \over 2 \, \delta x^* }
\geq \sqrt{ {\hbar T_{obs} \over 2}
\left( {1 \over  M_b} + {1 \over  M_d} \right)}
~,
\label{deltawignOLD}
\end{eqnarray}
where $M_b$ is the mass of 
the body, $M_d$ is the total mass of the device composed of
the light-gun, the detector, and the clock,
and the right-hand-side relation
follows from observing that Heisenberg's {\it Uncertainty Principle} 
implies $\delta x^* \delta v^* \ge (1/M_b + 1/M_d) \hbar/2$.
[N.B.: the {\it reduced mass} $(1/M_b+1/M_d)^{-1}$ 
is relevant for the relative motion.]
Clearly, from (\ref{deltawignOLD}) it follows that 
in order to eliminate the contribution to the uncertainty
coming from the quantum properties of the devices
it is necessary to take
the formal ``classical-device limit,''
{\it i.e.} the limit\footnote{A rigorous 
definition of a ``classical device'' is 
beyond the scope of this Article. However, it should be emphasized 
that the experimental setups being here considered require
the devices to be accurately positioned during the time
needed for the measurement, and therefore an ideal/classical
device should be infinitely massive so 
that the experimentalists can prepare it in a state 
with $\delta x \, \delta v \sim \hbar/M \sim 0$.
It is the fact that the infinite-mass limit is not accessible
in a gravitational context that forces one to 
consider only ``non-classical devices.'' 
This observation is not inconsistent with
conventional analyses of decoherence for macroscopic systems;
in fact, in appropriate environments, the behavior of a macroscopic 
device will still be ``closer to classical'' than the behavior of
a microscopic device, although the limit in which a device has exactly 
classical behavior is no longer accessible.}
of infinitely large $M_d$.

Up to this point I have not yet taken into account the
gravitational properties of the devices and in fact
the ``classical-device limit'' encountered above
is fully consistent with the laws of ordinary
Quantum Mechanics.
From a physical/phenomenological and conceptual
viewpoint it is well understood that the formalism
of Quantum Mechanics is only appropriate for the
description of the results of measurements performed
by classical devices. It is therefore not surprising
that the classical-device (infinite-mass) limit turned
out to be required 
in order to reproduce the prediction $min \delta D = 0$
of ordinary Quantum Mechanics
(which, as well known, allows $\delta A = 0$ for any
single observable $A$, since it only limits the combined
measurability of pairs of conjugate observables).

If one also takes into account
the gravitational properties of the devices,
a conflict with ordinary Quantum Mechanics
immediately arises because the
classical-device (infinite-mass) limit
is in principle inadmissible for measurements
concerning gravitational effects.\footnote{This conflict
between the
infinite-mass classical-device limit
(which is implicit in the applications of the formalism of ordinary
Quantum Mechanics to the description of the outcome of
experiments)
and the nature of gravitational interactions 
has not been addressed within any of the
most popular Quantum Gravity approaches,
including ``Canonical/Loop
Quantum Gravity''~\cite{canoloop} and ``Critical
Superstring Theory''~\cite{dbrane,critstring}.
In a sense somewhat similar to the one appropriate for
Hawking's work on black holes~\cite{hawkbh},
this ``classical-device paradox''
appears to provide an obstruction~\cite{gacgrf98} for the use
of the ordinary formalism of Quantum Mechanics
for a description of Quantum Gravity.}
As the devices get more and more massive they increasingly 
disturb the gravitational/geometrical observables, and
well before reaching the infinite-mass limit the procedures 
for the measurement of gravitational observables cannot
be meaningfully performed~\cite{gacmpla,gacgrf98,ng}.
In the Salecker-Wigner measurement procedure
the limit $M_d \rightarrow \infty$
is not admissible when gravitational interactions
are taken into account.
At the very least the value of $M_d$ is limited by the
requirement that the apparatus should not turn into a black hole
(which would not allow the exchange of signals
required by the measurement procedure).
These observations, which render unavoidable 
the $\sqrt{T_{obs}}$-dependence of Eq.~(\ref{deltawignOLD}),
provide motivation for the possibility~\cite{gacmpla,gacgrf98}
that in Quantum Gravity any measurement that monitors
a distance $D$ for a time $T_{obs}$ is affected by quantum
fluctuations such that\footnote{Note
that Eq.(\ref{deltagacdl}) sets a minimum uncertainty
which takes only into account the 
quantum and gravitational properties of the measuring
apparatus.
Of course, an even tighter
bound might emerge when taking
into account also the quantum and gravitational properties of
the system under observation. However, according
to the estimates provided in Refs.~\cite{padma,garay}
the contribution to the uncertainty coming from the system
if of the type $\delta D  \ge L_{planck}$,
so that the total contribution (summing the system
and the apparatus contributions) would be
of the type $\delta D  \ge L_{planck}
+ \sqrt{ {L_{QG} \, c \, T_{obs}}}$ which in nearly all contexts one
can be concerned with (which would have $c \, T_{obs} \gg L_{planck}$
can be approximated by completely neglecting the $L_{planck}$ correction
originating from the 
quantum and gravitational properties of the system.}
\begin{eqnarray}
\delta D  \ge \sqrt{ {L_{QG} \, c \, T_{obs}}}
~,
\label{deltagacdl}
\end{eqnarray}
where $L_{QG}$ could in principle be an independent
fundamental length scale (a length scale characterizing
the nature of the novel Quantum-Gravity
relationship between system and apparatus),
but one is tempted to consider the possibility
that $L_{QG}$ be simply related to the Planck
length.
Interestingly, according to (\ref{deltagacdl})
the Salecker-Wigner measurement of a distance $D$, which
requires a time $2 D/c$, would be affected
by an uncertainty of magnitude $\sqrt{L_{QG} D}$.

A $\delta D$ that increases with $T_{obs}$ 
({\it e.g.} as in (\ref{deltagacdl}))
is not surprising
for space-time fuzziness scenarios; in fact, the same phenomena
that would lead to fuzziness are also expected to 
induce ``information loss'' \cite{hawk}
(the information stored in a quantum system
degrades as $T_{obs}$ increases).
The argument based on the Salecker-Wigner setup provides motivation
to explore the specific form $\delta D \! \sim \! \sqrt{T_{obs}}$
of this $T_{obs}$-dependence.

Of course, the analyses reported above and
in Ref.~\cite{gacmpla,gacgrf98} do not necessarily indicate
that fuzziness of the type operatively defined in Section 2
should be responsible for the measurability bound (\ref{deltagacdl}).
The intuitive/heuristic arguments I advocated
can provide a (tentative) estimate of the measurability bound,
but a full Quantum Gravity theory woud be required in order
to be able to determine which phenomena could be
responsible for the bound. If one assumes that indeed
fuzziness of the type operatively defined in Subsection 2
is responsible for the measurability bound (\ref{deltagacdl})
one is led to the possibility that 
a distance $D$ would be affected
by fundamental stochastic fluctuations with
root-mean-square deviation $\sigma_D$ given by
\begin{eqnarray}
\sigma_D  \sim \sqrt{ {L_{QG} \, c \, T_{obs}}}
~.
\label{gacdl}
\end{eqnarray}
 
From the type of $T_{obs}$-dependence of Eq.~(\ref{gacdl})
it follows that the quantum fluctuations responsible
for (\ref{gacdl}) should have
displacement amplitude spectral density $S(f)$
with the $f^{-1}$
dependence\footnote{Of course, one expects that
an $f^{-1}$ dependence of the Quantum-Gravity induced $S(f)$
could only be valid for frequencies $f$ significantly
smaller than the Planck frequency $c/L_{planck}$
and significantly larger than the inverse of the time scale
over which, even ignoring the gravitational field
generated by the devices, the classical geometry 
of the space-time region where
the experiment is performed manifests
significant curvature effects.}
typical of ``random walk noise''~\cite{rwold}:
\begin{eqnarray}
S(f) = f^{-1} \sqrt {L_{QG} \, c} ~.
\label{gacspectr}
\end{eqnarray}
In fact, there is a general relation
(which follows~\cite{rwold} from the
general property (\ref{gacspectrule}))
between $\sigma_D  \sim \sqrt{T_{obs}}$
and $S(f) \sim f^{-1}$. 

If indeed $L_{QG} \sim L_{planck}$, from (\ref{gacspectr})
one obtains $S(f) \sim f^{-1} \cdot (5 \cdot 10^{-14} m \sqrt{H\!z})$.
As I shall discuss in detail later, by the standards of modern
interferometers this noise level is quite significant, and therefore,
before discussing other estimates of distance fuzziness,
let us see whether the naive guess $L_{QG} \sim L_{planck}$
can be justified within
the argument used in arriving at (\ref{deltagacdl}).
Since (\ref{deltagacdl}) was motivated from (\ref{deltawignOLD}),
and in going from (\ref{deltawignOLD}) to (\ref{deltagacdl})
the scale $L_{QG}$ was introduced to parametrize the
minimum allowed value of $1/M_b + 1/ M_d$,
we could get some intuition for $L_{QG}$
from trying to establish this minimum allowed value
of $1/M_b + 1/ M_d$.
As mentioned, a conservative 
(possibly very conservative)
estimate of this minimum value can be obtained by
enforcing that $M_b$ and $M_d$ 
be at least sufficiently small to avoid black hole formation.
In leading order ({\it e.g.}, assuming corresponding
spherical symmetries) this amounts to the requirement
that $M_b < \hbar S_b/(c L_{planck}^2)$
and  $M_d < \hbar S_d/(c L_{planck}^2)$,
where the lengths $S_b$ and $S_d$ characterize
the {\it sizes} of the regions of space where
the matter distributions associated to $M_b$ and $M_d$
are localized.
This observation implies
\begin{eqnarray}
{1 \over M_b} + {1 \over M_d}  >
{c L_{planck}^2 \over \hbar}
\left( {1 \over S_b} + {1 \over S_d} \right) 
~.
\label{estimatea}
\end{eqnarray}
This suggests~\cite{gacmpla}
that $L_{QG} \sim min [L_{planck}^2 (1/S_b + 1/S_d)]$:
\begin{eqnarray}
\delta D  \ge min \sqrt{ {{ \left( {1 \over S_b} 
+ {1 \over S_d} \right)
{L_{planck}^2  \, c \, T_{obs} \over 2}}}}
~.
\label{deltagacdlbis}
\end{eqnarray}
Of course, this estimate is very preliminary
since a full Quantum Gravity would be needed here; in particular,
the way in which black holes were handled
in my argument might have missed important properties
which would become clear only once we have the correct theory.
However, it is nevertheless striking to observe that the
naive guess $L_{QG} \sim L_{planck}$ appears extremely far from
the intuition emerging from this estimate; in fact,
$L_{QG} \sim L_{planck}$ would require that the maximum
admissible value
of $S_d$ be of order $L_{planck}$. [I take $S_b$ as fixed since
it characterizes the size of the bodies whose distance
is being measured, but of course the observer can choose
the size $S_d$ of the devices.]
Since our analysis only holds for bodies and devices that can
be treated as approximately rigid\footnote{The fact that
I have included only one contribution
from the quantum properties of the devices, the one associated
to the quantum properties of the motion of the center of mass,
implicitly relies on the assumption that the devices and the bodies
can be treated as approximately rigid. Any non-rigidity of the devices
would of course introduce additional contributions to the uncertainty
in the measurement of $D$. I shall further comment on the
additional uncertainties that are
introduced by the non-rigidity of devices in Section~5, where I 
consider some properties
of the mirrors used in gravity-wave interferometry.}
and any non-rigidity would introduce additional
contributions to the uncertainties, it is reasonable to assume
that $max[S_{d}]$ be some small length (small enough that
any non-rigidity would negligibly affect the measurement procedure),
but the condition $max[S_{d}] \sim L_{planck}$ appears rather extreme.
As I shall discuss in Section~4,
already available experimental data rule out $L_{QG} \sim L_{planck}$
in Eq.~(\ref{gacspectr}), and therefore if the $f^{-1}$-dependence
of Eq.~(\ref{gacspectr}) is verified in
the physical world (which is of course only
one of the possibilities, and a rather speculative one) $max[S_{d}]$
must be somewhat larger than $L_{planck}$.
As long as this type of analysis involves a $max[S_{d}]$
which is independent of $\delta D$ one still
finds $\sqrt{T_{obs}}$-dependence of $\sigma_D$
({\it i.e.} $f^{-1}$-dependence of $S(f)$).
If the correct Quantum Gravity is such that something
like (\ref{deltagacdlbis}) holds but with $max[S_{d}]$
that depends on $\delta D$, one would have a
different $T_{obs}$-dependence 
(and corresponding $f$-dependence), as I shall
show in one example discussed in Subsection~3.6.

\subsection{Random-walk noise from random-walk models
of quantum space-time fluctuations}

Since in this Article, like in Ref.~\cite{gacgwi},
I am advocating a rather pragmatic phenomenological approach
to Quantum Gravity, and taking into account the operative
definition of fuzzy distance given in Section~2,
it seems reasonable to consider the possibility that
the properties of a distance $D$ in a quantum space-time would involve
a fluctuation of magnitude $L_{planck} \sim 10^{-35} m$
over each time interval $t_{planck} = L_{planck}/c \sim 10^{-44} s$.
The type of interferometer noise that would
result from such a random-walk model
of quantum space-time has the same qualitative structure
as the noise I discussed in the previous Subsection motivated by the 
Salecker-Wigner measurement procedure.
In fact, experiments monitoring
the distance $D$ between two bodies for a time $T_{obs}$
(in the sense appropriate, {\it e.g.}, for a gravity-wave
interferometer)
would involve a total effect associated to quantum space-time
amounting to $n_{obs} \equiv T_{obs}/t_{planck}$ 
randomly directed fluctuations of magnitude $L_{planck}$.
An elementary analysis
allows to establish that in such a context
the root-mean-square deviation $\sigma_D$ would be
proportional to $\sqrt{T_{obs}}$:
\begin{equation}
\sigma_D \sim  \sqrt{L_{planck} c T_{obs}} \, .
\label{no1bis}
\end{equation}
We encounter again the $\sqrt{T_{obs}}$-dependence
already considered in relation to the analysis of the 
Salecker-Wigner measurement procedure.
Of course, this means that also for this random-walk models
of quantum space-time the
displacement amplitude spectral density has the
characteristic $f^{-1}$ behaviour.
It also means that Eq.~(\ref{no1bis}) as it stands
predicts too much fuzziness.
Therefore, if such a random-walk model of quantum space-time 
is verified in the physical world it must be that some of the simplifying
assumptions made in deriving Eq.~(\ref{no1bis}) were too naive.
One possibility one might want to consider is the one in
which the quantum properties of space-time 
are such that 
fluctuations of magnitude $L_{planck}$
would occur with frequency
somewhat lower than $1/t_{planck}$.

In closing this Subsection it seems worth adding a few comments
on the stochastic processes here considered.
In most physical contexts a series of random steps does not
lead to $\sqrt{T_{obs}}$ dependence of $\sigma$
because often the
context is such that through the
fluctuation-dissipation theorem
the source of $\sqrt{T_{obs}}$ dependence gets tempered.
The hypothesis explored in this Subsection,
which can be partly motivated from the analysis of
the Salecker-Wigner measurement procedure reported in
the previous Subsection,
is that the type of underlying dynamics of quantum space-time
be such that the fluctuation-dissipation theorem be satisfied
without spoiling the $\sqrt{T_{obs}}$ dependence of $\sigma$.
This is an intuition which apparently is shared by other
authors; in fact, the study reported in Ref.~\cite{garayclock}
(which followed by a few months Ref.~\cite{gacgwi},
but clearly was the result of completely independent work)
also models some implication of quantum space-time
(the ones that affect clocks) with stochastic processes
whose underlying dynamics does not produce any dissipation
and therefore the ``fluctuation contribution'' to
the $T_{obs}$ dependence remains unaffected,
although the fluctuation-dissipation theorem
is fully taken into account.

Since the mirrors of interferometers are basically extremities
of a pendulum, another aspect that the reader might at first
find counter-intuitive is that the $\sqrt{T_{obs}}$ dependence
of $\sigma$, although coming in with a very small prefactor,
for extremely large $T_{obs}$ would seem to give values of $\sigma$
too large to be consistent
with the structure of a pendulum.
This is a misleading intuition which originates from the experience
with ordinary (non-Quantum-Gravity) analyses of the pendulum.
In fact, the dynamics of an ordinary pendulum
has one extremity ``fixed'' to a very heavy and rigid body,
while the other extremity is fixed to a much lighter body.
The usual stochastic processes considered in
the study of the pendulum affect the heavier body
in a totally negligible way, while they have strong impact
on the dynamics of the lighter body.
A pendulum analyzed in the spirit of the present Subsection
would be affected by stochastic processes which are of the
same magnitude both for its heavier and its lighter extremity.
In particular in the directions orthogonal to
the vertical axis the stochastic processes affect the
position of the center of mass of the entire pendulum
just as they would affect the position of the center of
mass of any other body (the string that connects the two
extremities of the pendulum would not affect the motion of its
center of mass).

\subsection{Random-walk noise motivated by linear deformation 
of dispersion relation}

Both the analysis of the Salecker-Wigner measurement procedure
and the analysis of simple-minded 
random-walk models of quantum space-time fluctuations
have provided some encouragement
for the study of interferometer noise of random-walk type.
A third candidate Quantum Gravity effect
that provides some
encouragement for the random-walk noise scenario
has emerged in the context of 
studies~\cite{grbgac,aemn1,gampul,kpoin,lukipap}
of Quantum-Gravity induced deformation of the dispersion
relation that characterizes the propagation of massless particles.

Deformed dispersion relations are not uncommon
in the Quantum Gravity literature. For example, they emerge
naturally in Quantum Gravity scenarios requiring a modification
of Lorentz symmetry. Modifications of 
Lorentz symmetry could result from space-time
discreteness,
a possibility extensively investigated in the Quantum Gravity literature
(see, {\it e.g.}, Ref.~\cite{thooft}), 
and it would also naturally result from an ``active'' Quantum-Gravity
vacuum of the type advocated by Wheeler and Hawking~\cite{wheely,hawk}
(such a vacuum might physically label the space-time points).

While most Quantum-Gravity approaches will lead
to deformed dispersion 
relations, the specific structure of the deformation
can differ significantly from model to model.
Assuming that the deformation admits 
a series expansion at small energies $E$, and parametrizing 
the deformation in terms of an energy\footnote{I 
parametrize deformations of dispersion relations
in terms of an energy scale $E_{QG}$, which is implicitly 
assumed to be rather close to $E_{planck}$, while I parametrize
the proposals for measurability bounds
with a length scale $L_{QG}$,
which is implicitly 
assumed to be rather close to $L_{planck}$.
This is somewhat redundant, since of 
course $E_{planck} = \hbar c/L_{planck}$,
but it can help the reader in identifying the origin of a
conjectured fuzziness scenario by simply looking at the
type of parametrization that describes the stochastic processes.}
scale $E_{QG}$
(a scale characterizing the onset of
Quantum-Gravity dispersion effects, often identified with
the Planck energy $E_{planck} \sim 10^{19} GeV$),
one would expect to be able to approximate
the deformed dispersion relation at low energies according to
\begin{equation}
c^2{\bf p}^2 \simeq E^2 \left[1 
+ \xi \left({E \over E_{QG}}\right)^{\alpha}
\right]
\label{dispgen}
\end{equation}
where the power $\alpha$ and the  sign ambiguity $\xi = \pm 1$ 
would be fixed in a given dynamical framework.
For example, in some of the approaches
based on dimensionful ``$\kappa$'' quantum deformations of Poincar\'e
symmetries~\cite{kpoin,lukipap}
one finds evidence of a dispersion relation for
massless particles $c^2{\bf p}^2 = E_{QG}^2 \, \left[ 1 -
e^{E/E_{QG}}\right]^2$, and therefore $\xi = \alpha = 1$.

Scenarios (\ref{dispgen}) with $\alpha = 1$ are in a sense
consistent with random-walk noise. 
In fact, an experiment involving as
a device (as a probe) a massless particle satisfying the
dispersion relation (\ref{dispgen}) with $\alpha = 1$
would be naturally
affected by a device-induced uncertainty that grows
with $\sqrt{T_{obs}}$. 
This is for example true in Quantum-Gravity scenarios in
which the Hamiltonian equation of 
motion ${\dot x}_i\,=\partial\,H/\partial\,p_i$ is still valid (at least
approximately), where 
the deformed dispersion relation (\ref{dispgen})
leads to energy-dependent velocities
for massless particles\cite{grbgac,aemn1,kpoin,lukipap} 
\begin{equation}
v \simeq c \left[1 
- \left( {1+\alpha \over 2} \right)
 \xi \left({E \over E_{QG}}\right)^{\alpha} \right] ~,
\label{velogen}
\end{equation}
and consequently the uncertainty in the position of the massless
probe when a time $T_{obs}$ has lapsed since the observer
(experimentalist) set off the measurement procedure
is given by
\begin{equation}
\delta x \simeq  c \, \delta t + \delta v \, T_{obs} 
\simeq c \, \delta t + {1+\alpha \over 2} \, \alpha  \,
{E^{\alpha-1} \, \delta E \over E_{QG}^{\alpha}} c \, T_{obs} ~,
\label{deltagacgen}
\end{equation}
where $\delta t$ is the quantum uncertainty
in the time of emission of the probe, 
$\delta v$ is the quantum uncertainty
in the velocity of the probe,
$\delta E$
is the quantum uncertainty
in the energy of the probe,
and I used the relation
between $\delta v$ and $\delta E$ that follows from (\ref{velogen}).
Since the quantum uncertainty
in the time of emission of a particle and the quantum uncertainty
in its energy are related\footnote{It is well understood
that the $\delta t \, \delta E \ge \hbar$ relation
is valid only in a weaker sense than, say,
Heisenberg's Uncertainty Principle $\delta x \, \delta p \ge \hbar$.
This has its roots in the fact that the time appearing in
Quantum-Mechanics equations is just a parameter (not an operator),
and in general there is no self-adjoint operator canonically conjugate
to the total energy, if the energy spectrum is bounded 
from below~\cite{pauli,garayclock}.
However, the $\delta t \, \delta E \ge \hbar$ relation
does relate $\delta t$ intended as quantum uncertainty
in the time of emission of a particle and $\delta E$
intended as quantum uncertainty
in the energy of that same particle.}
by $\delta t \, \delta E \ge \hbar$, Eq.~(\ref{deltagacgen})
can be turned into an absolute bound on 
the uncertainty in the position of the massless
probe when a time $T_{obs}$ has lapsed since the observer
set off the measurement procedure:
\begin{equation}
\delta x \ge c  {\hbar \over \delta E} 
+ {1+\alpha \over 2} \, \alpha  \,
{E^{\alpha-1} \, \delta E \over E_{QG}^{\alpha}} T_{obs} 
\ge \sqrt{\left({\alpha+\alpha^2 \over 2} \right)
\left( {E \over E_{QG}}\right)^{\alpha-1}
{c^2 \hbar T_{obs} \over E_{QG}}}
~, 
\label{deltagacgenfin}
\end{equation}
where I also used the fact that in principle the observer
can prepare the probe in a state with desired $\delta t$,
so it is legitimate to minimize the uncertainty with respect
to the free choice of $\delta t$.

For $\alpha=1$ the $E$-dependence on the right-hand side of 
Eq.~(\ref{deltagacgenfin}) disappears and one is led again
(see Subsections 3.2 and 3.3) to 
a $\delta x$ of the type $(constant) \cdot \sqrt{T_{obs}}$:
\begin{equation}
\delta x \ge \sqrt{{c^2 \hbar T_{obs} \over E_{QG}}}
~.
\label{deltagacgenfinalphaone}
\end{equation}

When massless probes are used in the measurement of a distance $D$,
as in the Salecker-Wigner measurement procedure,
the uncertainty (\ref{deltagacgenfinalphaone}) in the position
of the probe translates directly into an uncertainty on $D$:
\begin{equation}
\delta D \ge \sqrt{{c^2 \hbar T_{obs} \over E_{QG}}}
~.
\label{deltagacgenfinalphaoned}
\end{equation}
This was already observed in Refs.~\cite{gacxt,aemn1,lukipap}
which considered the implications of deformed dispersion 
relations (\ref{dispgen}) with $\alpha = 1$ 
for the Salecker-Wigner measurement procedure.

Since deformed dispersion 
relations (\ref{dispgen}) with $\alpha = 1$ 
have led us to the same measurability bound 
already encountered both in the analysis of the Salecker-Wigner 
measurement procedure and the analysis of simple-minded 
random-walk models of quantum space-time fluctuations,
if we assume again that such measurability bounds
emerge in a full Quantum Gravity as a result of
corresponding quantum fluctuations (fuzziness), we are led
once again to random-walk noise:
\begin{equation}
\sigma_D \sim \sqrt{{c^2 \hbar T_{obs} \over E_{QG}}}
~.
\label{deltagacgenfinalphaonesigmad}
\end{equation}

\subsection{Noise motivated by quadratic deformation 
of dispersion relation}

In the preceding Subsection~3.4 I observed that
Quantum-Gravity deformed dispersion 
relations (\ref{dispgen}) with $\alpha = 1$ 
can also motivate random-walk 
noise $\sigma_D \sim (constant) \cdot \sqrt{T_{obs}}$.
If we use the same line of reasoning that connects a
measurability bound to a scenario for fuzziness
when $\alpha \ne 1$
we find $\sigma_D \sim c(E/E_{QG}) \cdot \sqrt{T_{obs}}$,
where $c(E/E_{QG})$ is a ($\alpha$-dependent) function
of $E/E_{QG}$. However, in these cases with $\alpha \ne 1$
clearly the connection between measurability bound
and fuzzy-distance scenario cannot be too direct; in fact,
the energy of the probe $E$ which naturally playes a role in
the context of the derivation of the measurability bound
does not have a natural counter-part 
in the context of the conjectured fuzzy-distance scenario.

In order to preserve the 
conjectured connection between measurability bounds
and fuzzy-distance scenarios one can be tempted to envision that
if $\alpha \ne 1$ the interferometer noise levels induced
by space-time fuzziness might be of the type
[see Eq.~(\ref{deltagacgenfin})]
\begin{equation}
\sigma_D \sim \sqrt{\left({\alpha+\alpha^2 \over 2} \right)
\left( {E^* \over E_{QG}}\right)^{\alpha-1}
{c^2 \hbar T_{obs} \over E_{QG}}} ~,
\label{deltagacgenfinalphamanysigmad}
\end{equation}
where $E^*$ is some energy scale characterizing the physical context
under consideration. [For example, at the intuitive level
one might conjecture that $E^*$
could characterize some sort of energy
density associated with quantum fluctuations of space-time
or an energy scale associated with the masses of the
devices used in the measurement process.]

Since $\alpha \ge 1$ in all Quantum-Gravity approaches 
believed to support deformed dispersion relations,
and since it is quite plausible that $E_{QG}$ would be 
rather close to $10^{19} GeV$, it appears likely that
the factor $(E^*/E_{QG})^{\alpha -1}$ would suppress the 
random-walk noise effect.

\subsection{Noise with $f^{-5/6}$ amplitude spectral density}

In Subsection~3.2 a 
bound on the measurability of distances based on the 
Salecker-Wigner procedure was used as motivation
for experimental tests of 
interferometer noise of
random-walk type, with $f^{-1}$ amplitude spectral density
and $\sqrt{T_{obs}}$ root-mean-square deviation.
In this Subsection I shall pursue further the observation
that the relevant measurability bound could be derived
by simply insisting that the devices do not turn into black holes.
That observation allowed to derive Eq.~(\ref{deltagacdlbis}),
which expresses the minimum uncertainty $\delta D$
on the measurement of a distance $D$ ({\it i.e.} the measurability
bound for $D$) as proportional to $\sqrt{T_{obs}}$ and 
$\sqrt{(1/S_b +1/ S_d)}$.
Within that derivation
the minimum uncertainty is therefore obtained 
in correspondence of the minimum value of $1/S_b+1/S_d$
consistent with the structure of the measurement procedure.
Since, given the size $S_b$ of the bodies whose distance
is being measured, the minimum of $1/S_b + 1/S_d$ corresponds 
to $max[S_{d}]$ I was led to consider how large $S_d$ could be 
while still allowing to disregard any non-rigidity in the quantum
motion of the device (which would otherwise lead to additional
contributions to the uncertainties).
I managed to motivate the random-walk
noise scenario by simply assuming that $max[S_{d}]$ be independent of 
the accuracy $\delta D$ that the observer would wish to achieve.
However, as already argued earlier in this Article,
the same physical intuition that motivates some of the fuzzy space-time
scenarios here considered 
also suggests that Quantum Gravity might require a novel measurement
theory, possibly involving a new 
type of relation between system and measuring apparatus.
Based on this intuition,
it seems reasonable to contemplate the possibility
that $max[S_{d}]$  might actually depend on $\delta D$.

It is such a scenario that I want to consider in this Subsection.
In particular I want to consider the case $max[S_{d}] \sim \delta D$,
which, besides being simple, has the plausible
property that it allows only small devices if the uncertainty
to be achieved is small, while it would allow 
correspondingly larger devices if the observer was
content with a larger uncertainty.
This is also consistent with the idea that elements of non-rigidity
in the quantum motion of extended devices might be negligible
if anyway the measurement is not aiming for great accuracy,
while they might even lead to the most 
significant contributions to the uncertainty if all other sources 
of uncertainty are very small.
Salecker and Wigner~\cite{wign} would 
also argue that ``large'' devices
are not suitable for very accurate space-time measurements
(they end up being ``in the way'' of the measurement procedure)
while they might be admissible if space-time is being probed 
rather softly.

In this scenario with $max[S_{d}] \sim \delta D$,
Eq.~(\ref{deltagacdlbis}) takes the form
\begin{eqnarray}
\delta D  \ge \sqrt{ {{ \left( {1 \over S_b} 
+ {1 \over S_d} \right)
{L_{planck}^2  \, c \, T_{obs} \over 2}}}}
\ge \sqrt{ {{ L_{planck}^2 \, c \, T_{obs} \over 2\,\, \delta D }}}
~,
\label{predeltagactwothird}
\end{eqnarray}
which actually gives
\begin{eqnarray}
\delta D  \ge \left( {{1 \over 2} L_{planck}^2 \, 
c \, T_{obs} }\right)^{1/3}
~.
\label{deltagactwothird}
\end{eqnarray}
As already done with 
the other measurability bounds discussed in this Article,
I shall take Eq.~(\ref{deltagactwothird}) as motivation
for the investigation of 
the corresponding fuzziness scenarios characterised by
\begin{eqnarray}
\sigma_D  \sim \left( {\tilde L}_{QG}^2 \, 
c \, T_{obs} \right)^{1/3}
~.
\label{sigmagactwothird}
\end{eqnarray}
Notice that in this equation I replaced $L_{planck}$
with a generic length scale ${\tilde L}_{QG}$, 
since it is possible that the heuristic
argument leading to Eq.~(\ref{sigmagactwothird}) 
might have captured
the qualitative structure of the phenomenon while providing 
an incorrect estimate of the relevant length scale.
As discussed later in this Article significant
bounds on this length scale can be set by experimental data,
so we can take a phenomenological attitude toward ${\tilde L}_{QG}$.

As one can verify for example using Eq.~(\ref{gacspectrule}),
the $T_{obs}^{1/3}$ dependence of $\sigma_D$ is associated with
displacement amplitude spectral density with $f^{-5/6}$ behaviour:
\begin{eqnarray}
{\cal S}(f) = f^{-5/6} ({\tilde L}_{QG}^2 \, c)^{1/3}
~.
\label{no3nspectr}
\end{eqnarray}
For ${\tilde L}_{QG} \sim 10^{-35} m$ this equation would 
predict ${\cal S}(f) = f^{-5/6} \cdot (3 \cdot 10^{-21} m \, H\!z^{1/3})$.

\section{COMPARISON WITH GRAVITY-WAVE $~~~~~~~~~~~~~~~~$
INTERFEROMETER DATA}

From the point of view of the operative definition
of fuzzy distance given in Section~2 the scenarios 
for space-time fuzziness considered
in the previous Section can all be characterized in
terms of three alternative possibilities for the
root-mean-square deviation $\sigma_D$ associated to 
the fluctuations induced on $D$ by conjectured 
quantum properties of space-time.
For convenience I report here the three 
alternatives for $\sigma_D$ that I ended up considering:
\begin{equation}
\sigma_D \sim L_{min} \, ,
\label{no1n}
\end{equation}
\begin{eqnarray}
\sigma_D  \sim \sqrt{ {L_{QG} \, c \, T_{obs}}}
~,
\label{no2n}
\end{eqnarray}
\begin{eqnarray}
\sigma_D  \sim \left( {\tilde L}_{QG}^2 \, 
c \, T_{obs} \right)^{1/3}
~.
\label{no3n}
\end{eqnarray}

The discussion of the fuzziness scenarios considered
in the previous Section was consistent with the assumption
that the length scale characterizing fuzziness
(be it $L_{min}$, $L_{QG}$ or ${\tilde L}_{QG}$)
would be a general fundamental property of Quantum Gravity,
independent of the 
peculiarities of the specific experimental setup
and of its environment. However, the fuzziness scenario
considered in Subsection~3.5 provided some motivation
for the idea that at least $L_{QG}$ (if Eq.~(\ref{no2n}) was
to be realized in the physical world)
might not be a universal length scale,
{\it i.e.} it might depend on some specific properties of the 
experimental setup and in particular in some contexts
(those with small $E^*/E_{QG}$) one might find ${\tilde L}_{QG}$
to be significantly smaller than $L_{planck}$.
The possibility that the ``magnitude'' 
of space-time fuzziness might depend on the specific context
and experimental setup is also consistent with the arguments
which support the possibility of a novel Quantum-Gravity
relationship between system and measuring apparatus.
If the length scale characterizing fuzziness
depended on this relationship it might take different
values in different experimental setups.

Setting aside these possible complications associated to
a novel Quantum-Gravity
relationship between system and measuring apparatus,
I shall proceed discussing the bounds set on
the length scales $L_{min}$, $L_{QG}$ or ${\tilde L}_{QG}$
by available experimental data.
Let me start observing that, while conceptually
they represent drastic departures from conventional physics, 
phenomenologically
the proposals (\ref{no1n}), (\ref{no2n}) and (\ref{no3n})
appear to encode only minute effects.
For example, assuming
that $L_{min}$, $L_{QG}$ and ${\tilde L}_{QG}$  are not much 
larger than the Planck length,
all of these proposals encode submeter uncertainties on 
the size of the whole observable universe
(about $10^{10}$ light years).
However, the precision~\cite{saulson}
of modern gravity-wave interferometers is such that
they can provide significant information at least on 
the proposals (\ref{no2n}) and (\ref{no3n}).
In fact, as already mentioned in Section~2,
the operation of gravity-wave interferometers 
is based on the detection of
minute changes in the positions of some test masses (relative to
the position of a beam splitter).
If these positions were 
affected by quantum fluctuations of the type discussed
above the operation of gravity-wave interferometers would effectively
involve an additional
source of noise due to Quantum-Gravity.
This observation allows to set interesting bounds already
using existing noise-level data obtained at
the {\it Caltech 40-meter interferometer}, which
has achieved~\cite{ligoprototype}
displacement noise levels with amplitude spectral density
lower than $10^{-18} m/\sqrt{H\!z}$
for frequencies between $200$ and $2000$ $H\!z$.
While these sensitivity levels are still very far from the levels
required in order to test proposal (\ref{no1n})
(from the analysis reported in Subsection~3.1
it follows that for $L_{min} \sim L_{planck}$
and $f \sim 1000 Hz$ the Quantum-Gravity noise induced in that
scenario is only of order $10^{-36} m/\sqrt{H\!z}$), 
as seen by straightforward comparison
with Eq.~(\ref{gacspectr})
these sensitivity levels 
clearly rule out all values of $L_{QG}$
down to the Planck length.
Actually, even values of $L_{QG}$ significantly
smaller than the Planck length are inconsistent with the data 
reported in Ref.~\cite{ligoprototype}; in particular,
by confronting Eq.~(\ref{gacspectr}) with the observed
noise level of $3 \cdot 10^{-19} m/\sqrt{H\!z}$ near $450$ $H\!z$,
which is the best achieved at the {\it Caltech 40-meter interferometer},
one obtains the bound $L_{QG} \le 10^{-40}m$.

While, as mentioned, 
at present we should allow for some relatively small factor
to intervene in the relation between $L_{QG}$ and $L_{planck}$,
the exclusion of all values of $L_{QG}$
down to $10^{-40}m$ 
appears to be quite significant,
perhaps even problematic, for the proposal (\ref{no2n}).
In particular, 
this experimental bound rules out the possibility that
(\ref{no2n}) might be the result of
space-time fluctuations of the
random-walk type discussed in
Subsection~3.3, with
a fluctuation of magnitude $L_{planck} \sim 10^{-35} m$
for each time interval $t_{planck} \sim 10^{-44} s$;
in fact, as shown above, such a picture would 
lead to values of $L_{QG}$ not significantly
smaller than $L_{planck}$.
The fact that this picture is ruled out 
is perhaps the most striking lesson coming out
of available interferometer data.
Only a few years ago it might have seemed impossible
to test a scenario involving 
fluctuations of magnitude $L_{planck}$,
even if such fluctuations might have been quite
frequent (one each $t_{planck}$).
From the point of view of modeling 
quantum fluctuations of space-time 
our simple-minded random-walk model might
still be useful, but clearly some new element
must be introduced in order to temper the
associated fuzziness of space-time; for example,
as mentioned in closing Subsection~3.3, one might
consider the possibility that fluctuations of 
magnitude $L_{planck}$ would not be as frequent 
as $1/t_{planck}$.

In any case, of course, even more stringent bounds on $L_{QG}$
are within reach of the next LIGO/VIRGO~\cite{ligo,virgo}
generation of gravity-wave interferometers. 
It would seem that very little room for adjustments of the
random-walk noise scenario would remain available
if also LIGO
and VIRGO give negative results for what concerns
this scenario.
 
The sensitivity achieved at
the {\it Caltech 40-meter interferometer}
also sets a bound on the proposal
(\ref{sigmagactwothird})-(\ref{no3nspectr}).
By observing that Eq.~(\ref{no3nspectr}) would imply Quantum-Gravity
noise levels for gravity-wave interferometers of
order ${\tilde L}_{QG}^{2/3} \cdot (10 \, m^{1/3}/\sqrt{H\!z})$
at frequencies of a few hundred $H\!z$, one obtains
from the data reported in Ref.~\cite{ligoprototype}
that ${{\tilde L}_{QG}} \le 10^{-29} m$.
This bound is remarkably stringent in absolute terms, but is still
quite far from the range of values one ordinarily considers
as likely candidates for length scales appearing in Quantum Gravity.
A more significant bound on ${{\tilde L}_{QG}}$ 
should be obtained by the LIGO/VIRGO generation 
of gravity-wave interferometers.
For example, it is plausible~\cite{ligo} that 
the ``advanced phase'' of LIGO achieve a displacement noise spectrum 
of less than $10^{-20} m/ \sqrt{H\!z}$ near $100$ $H\!z$
and this would probe values of ${{\tilde L}_{QG}}$ as small
as $10^{-34} m$.
 
Looking beyond the LIGO/VIRGO generation
of gravity-wave interferometers, one can envisage still quite
sizeable margins for improvement by optimizing the
performance of the interferometers at low frequencies,
where both (\ref{gacspectr}) and (\ref{no3nspectr}) become
more significant. It appears natural to perform such studies
in the quiet environment of space, perhaps through future
refinements of LISA-type setups~\cite{lisa}.

The indication of the low-frequency range as most promising for
Quantum Gravity tests at interferometers
should be seen as the most robust result
obtained in this Article. The arguments advocated in the 
previous Section~3 were all rather speculative and it 
would not be surprising if some of the details of the estimates 
turned out to be completely off the mark, but the fact that nearly 
all of those arguments pointed us toward the low-frequency region
might nevertheless be indicative. 
I hope that,
in spite of the heuristic nature
of the arguments advocated in the previous Section,
colleagues
on the experimental side will
take the low-frequency hint into consideration
in planning future experimental tests of quantum properties
of space-time.
 
\section{ABSOLUTE MEASURABILITY BOUND FOR $~~~~~$
THE AMPLITUDE OF A GRAVITY WAVE}

Up to this point I have discussed how certain plausible
quantum properties
of space-time would affect the noise levels in interferometers.
The fact that I considered the sensitivity levels of gravity-wave 
interferometers
was due to the fact that these are the most advanced 
modern interferometers, and it was not in any way related 
to their function as gravity-wave detectors. 
In this Section~5 I instead consider 
an aspect of the physics of gravity waves; specifically, I discuss
the way in which the interplay 
between Gravity and Quantum Mechanics could affect the measurability
of the amplitude of a gravity wave.
The reader should notice that in this Section nothing is assumed
of Quantum Gravity: I just combine known properties of Gravity
and Quantum Mechanics.
This is also different from the analyses reported
in the previous sections which concerned candidate Quantum Gravity
phenomena. The motivation for considering those Quantum Gravity
phenomena
came from combining known properties of Gravity
and Quantum Mechanics, but the phenomena 
({\it e.g.} the models for space-time fuzziness)
could not be seen
as straightforward combination of Gravity
and Quantum Mechanics, they truly pertained to a novel type of physics.

Having clarified in which sense this Section represents a deviation
from the main bulk of observations reported in the present Article,
let me start the discussion
by reminding the reader of the fact that, as 
already mentioned in Section~2,
the interference pattern generated by a modern interferometer
can be remarkably sensitive to changes in the positions of the mirrors
relative to the beam splitter, and is therefore
sensitive to gravitational waves (which, as described in
the {\it proper reference frame}~\cite{saulson},
have the effect of changing these relative positions).
With just a few lines of simple algebra
one can show that an ideal gravitational wave
of amplitude $h$ and reduced\footnote{I report
these results
in terms of reduced wavelengths $\lambda^{o}$
(which are related to the wavelengths $\lambda$ by 
$\lambda^{o}= \lambda/(2 \pi)$) in order to avoid 
cumbersome factors of $\pi$ in some of the formulas.}
wavelength $\lambda^{o}_{gw}$
propagating along the direction orthogonal to
the plane of the interferometer would cause a change in the
interference pattern as for a phase
shift of magnitude $\Delta \phi = D_L/ \lambda^{o}$,
where $\lambda^{o}$ is the reduced
wavelength of the laser beam used in
the measurement procedure and~\cite{saulson,qigwdbook}
\begin{eqnarray}
D_L \sim 2 \, h \,  \lambda^{o}_{gw} \,
\left| \sin \left( {L \over
2 \lambda^{o}_{gw}} \right) \right|
~,
\label{dleq}
\end{eqnarray}
is the
magnitude of the change caused by the gravitational wave in the length 
of the arms of the interferometer. 
(The changes in the lengths of the two arms
have opposite sign~\cite{saulson}.)
 
As already mentioned in Section~2,
modern techniques allow to construct
gravity-wave interferometers with truly remarkable sensitivity;
in particular, at least for
gravitational waves with $\lambda^{o}_{gw}$ of
order $10^3 Km$, the next LIGO/VIRGO generation of detectors
should be sensitive to $h$ as low as $3 \cdot 10^{-22}$.
Since $h \sim 3 \cdot 10^{-22}$ causes a $D_L$ of order $10^{-18}m$
in arms lengths $L$ of order $3 Km$, it is not surprising 
that in the analysis of gravity-wave interferometers,
in spite of their huge size,
one ends up having to take into account~\cite{saulson}
the type of quantum effects usually significant only for the study of
processes at or below the atomic scale.
In particular, there is the so-called {\it standard quantum limit}
on the measurability of $h$ that results 
from the combined minimization
of {\it photon shot noise} and {\it radiation pressure noise}.
While a careful discussion of these two noise sources 
(which the interested
reader can find in Ref.~\cite{saulson})
is quite insightful, here I shall rederive
this {\it standard quantum limit} in an alternative\footnote{While
the {\it standard quantum limit} can be equivalently obtained
either from the combined minimization
of {\it photon shot noise} and {\it radiation pressure noise}
or from the application of Heisenberg's uncertainty
principle to the position and momentum of the mirror,
it is this author's opinion that there might
actually be a fundamental
difference between the two derivations.
In fact, it appears (see, {\it e.g.}, Ref.~\cite{jare1}
and references therein)
that the limit obtained through combined minimization
of {\it photon shot noise} and {\it radiation pressure noise}
can be violated by careful exploitation of the properties
of squeezed light, whereas the 
limit obtained through the application of Heisenberg's uncertainty
principle to the position and momentum of the mirror
is so fundamental that it could not possibly be violated.}
and straightforward manner (also discussed in Ref.~\cite{qigwdbook}),
which relies on the application of Heisenberg's uncertainty
principle to the position and momentum of a mirror relative
to the position of the beam splitter.
This can be done along the lines of my analysis
of the
Salecker-Wigner procedure for the measurement of distances.
Since the mirrors and the beam splitter 
are macroscopic, and therefore
the corresponding momenta and velocities are related 
non-relativistically, Heisenberg's uncertainty principle implies that
\begin{eqnarray}
\delta x \, \delta v \ge {\hbar \over 2} \left({1 \over M_m} 
+ {1 \over M_b} \right) \ge {\hbar \over 2 M_m}
~,
\label{deltaheis}
\end{eqnarray}
where $\delta x$ and $\delta v$ are the uncertainties in the
relative position and relative velocity,
$M_m$ is the mass of the mirror, $M_b$ is the mass
of the beam splitter. [Again, the relative motion is
characterised by the {\it reduced mass}, which is
given in this case by $(1/M_m+1/M_b)^{-1}$.] 

Clearly, the high precision of the planned measurements
requires~\cite{saulson,qigwdbook} that the
position of the mirrors
be kept under control during the whole time $2L/c$
that the beam spends in between the arms of
the detector before superposition.
When combined with (\ref{deltaheis}) this leads to the finding
that, for any given value of $M_m$,
the $D_L$ induced by the gravitational wave can be measured
only up to an irreducible uncertainty,
the so-called {\it standard quantum limit}:
\begin{eqnarray}
\delta D_L \ge  \delta x + \delta v \, 2 {L \over c}
\ge \delta x + {\hbar L \over c M_m \delta x} 
\ge  \sqrt{{\hbar L \over c M_m }}
~.
\label{deltawign}
\end{eqnarray}
Here of course the reader will realize that 
the conceptual steps are completely analogous to the
one of the discussion given in Section~3
of the Salecker-Wigner procedure for the measurement of distances.
The similarities between the analysis of measurability for
Salecker-Wigner distance measurements
and the analysis of measurability by 
gravity-wave interferometers 
are a consequence of the fact that in both contexts
a light signal is exchanged and the
measurement procedure requires that the relative positions of 
some devices be known with high accuracy during the whole time
that the signal spends between the bodies.

The case of gravity-wave measurements is a canonical
example of my general argument that the infinite-mass 
classical-device limit underlying ordinary Quantum Mechanics
is inconsistent with the nature of gravitational measurements.
As the devices get more and more massive they not only
increasingly disturb the gravitational/geometrical observables, 
but eventually 
(well before reaching the infinite-mass limit)
they also render impossible~\cite{gacmpla,gacgrf98} 
the completion of the procedure of measurement 
of gravitational observables.
In trying to asses how this observation affects
the measurability of the properties of a gravity wave
let me start by combining Eqs.~(\ref{dleq})
and (\ref{deltawign}):
\begin{eqnarray}
\delta h = \delta \left({D_L \over L}\right)
= h {\delta D_L \over D_L}
\ge {\sqrt{{\hbar L \over c M_m }} \over
2 \,  \lambda^{o}_{gw} \,
\left| \sin \left( {L \over 2 \lambda^{o}_{gw}} 
\right) \right| }
~.
\label{deltawignh}
\end{eqnarray}
In complete analogy with some of the observations
made in Section~3 concerning the measurability of distances,
I observe that, when  
gravitational effects are taken into account,
there is an obvious limitation on the mass of the mirror: $M_m$
must be small enough that
the mirror does not turn into a black hole.\footnote{This is of 
course a very conservative bound, since
a mirror stops being useful as a device well before it turns into
a black hole, but even this conservative approach leads 
to an interesting finding.}
In order for the mirror not to be a black hole
one requires $M_m < \hbar S_m/(c L_{planck}^2)$,
where $L_{planck} \sim 10^{-33} cm$ 
is the Planck length and $S_m$ is the {\it size} of 
the region of space occupied by the mirror.
This observation combined with (\ref{deltawignh})
implies that one would have obtained a bound on
the measurability of $h$
if one found a maximum allowed mirror size $S_m$.
In estimating this maximum $S_m$
one can be easily led to two extreme assumptions
that go in opposite directions.
It is perhaps worth commenting on the weaknesses
of these assumptions,
as this renders more intuitive the discussion
of the correct estimate.
On one extreme, one could suppose
that in order to achieve a sensitivity to $D_L$ as
low as $10^{-18} m$ it might be necessary to ``accurately position''
each $10^{-36} m^2$ surface element of the mirror.
If this was
really necessary,
our line of argument would then lead to a rather large
measurability bound.
Fortunately, 
the phase of the 
wavefront of the reflected light beam is determined
by the average position of all the atoms across the beam's width,
and microscopic irregularities in the structure of
the mirror only lead to scattering of a small fraction of light
out of the beam.
This suggests that in our analysis the size of the mirror
should be assumed to be of the order of the width of the 
beam~\cite{saulson}.
Once this is taken into account another extreme assumption
might appear to be viable. In fact, especially when guided
by intuition coming from table-top interferometers,
one might simply assume that the mirror could be {\it attached}
to a very massive body.
Within this assumption
our line of argument would not lead to any
bound on the measurability of $h$. However, whereas
for the type of accuracies typically involved in table-top
experiments the idealization of a mirror {\it attached} to the
table is appropriate\footnote{Of course,
in a table-top experiment it is possible to {\it attach} a mirror
to the table in such a way that the noise associated to the
residual relative motion of the mirror with respect to the table
be smaller than all other sources of noise.},
in gravity-wave interferometers 
the precision is so high that it becomes necessary to take
into account the fact that
no {\it attachment procedure} can violate Heisenberg's 
Uncertainty Principle
(and causality).
Clearly by {\it attaching}
a mirror of size $S_m$ to a massive body
one would not avoid
the minimum uncertainty $\sqrt{c T L_{planck}^2/S_m}$
in the position of the mirror over a time $T$.
Actually, mirrors 
provide a good context in which to illustrate
the interplay between gravitational and quantum
properties of devices.
If a mirror is extended enough that one might not be able to neglect
the fact it is not really
moving rigidly with its center of mass, additional contributions
to quantum uncertainties are found.
The relative motion of different parts of the mirror is, of course, 
not ``immune'' to the Uncertainty Principle, and
over a time $T_{obs}$ the relative position of different parts of the
mirror will necessarily have an uncertainty
proportional to $\sqrt{\hbar T_{obs}/m}$, where $m$ is the mass of
the small portions of the mirror whose relative position we are considering
(rather than the larger mass of the entire mirror).
In order to be able to use the mirror these uncertainties must be 
small enough to render the mirror consistent with the level of 
accuracy required by the
measurement.\footnote{It is worth emphasizing that ideal mirrors
(like other ideal classical devices)
are consistent with the laws of ordinary
(non-gravitational) Quantum Mechanics,
but are inadmissible once gravitational interactions are
turned on.
In the limit of ordinary Quantum Mechanics
in which each small portion of the mirror
has infinite mass the {\it Uncertainty Principle}
ceases to induce non-rigidity in the mirror.
However, when gravitational interactions combine with Quantum Mechanics
this infinite mass limit 
is inconsistent with the procedure of measurement of
gravitational observables.}
These considerations further supports the point that in 
the present analysis the size $S_m$ of the mirror
should be taken to be of the order of the width of the beam,
rather than being replaced by the size of some massive
body {\it attached} to the mirror.
In light of these considerations one clearly sees
an upper bound on $S_m$; in fact,
if the width of the beam 
(and therefore the effective size of the mirror) 
is larger than the $\lambda^{o}_{gw}$
of the gravity wave\footnote{Note that
for the gravitational waves to which LIGO will be most sensitive,
which have $\lambda^{o}_{gw}$ of
order $10^3 Km$, the
requirement $S_m < \lambda^{o}_{gw}$
simply states that the size of mirrors
should be smaller than $10^3 Km$.
This bound might appear very conservative, but I am
trying to establish an {\it in principle}
limitation on the measurability of $h$, and therefore
I should not take into account
that present-day technology is very far
from being able to produce a $10^3 Km$ mirror
with the required profile precision.} 
which one is planning to observe,
that same gravity wave would cause phenomena
that would not allow the proper completion of the measurement
procedure ({\it e.g.} deforming the mirror and leading to a nonlinear
relation between $D_L$ and $h$).
One concludes that $M_m$ should be smaller 
than $\hbar \lambda^{o}_{gw} /(c L_{planck}^2)$, and
this can be combined with (\ref{deltawignh})
to obtain the measurability bound 
\begin{eqnarray}
\delta h > {L_{planck} \over 2 \, \lambda^{o}_{gw}} 
{\sqrt{{ L/ \lambda^{o}_{gw} }} \over
\left| \sin \left( {L \over 2 \lambda^{o}_{gw}} 
\right) \right| }
~.
\label{hbound}
\end{eqnarray}
This result not only sets a lower bound on the measurability of $h$
with given arm's length $L$, but also
encodes an absolute ({\it i.e.}
irrespective of the value of $L$)
lower bound, as a result of the fact that
the function $\sqrt{x}/|\sin (x/2)|$
has an absolute minimum: $min[\sqrt{x}/\sin (x/2)] \sim 1.66$.
This novel measurability bound
is a significant departure from the principles of ordinary
Quantum Mechanics, especially in light of the fact that
it describes a limitation on the measurability of a single observable
(the amplitude $h$ of a gravity wave),
and that this limitation turns out to depend on the value (not the 
associated uncertainty) of another observable
(the reduced wavelength $\lambda^{o}_{gw}$
of the same gravity wave).
It is also significant that this new bound (\ref{hbound})
encodes an aspect of a novel type of interplay
between system and measuring apparatus in Quantum-Gravity regimes;
in fact, in deriving (\ref{hbound}) a crucial role was played by the
fact that in accurate measurements of gravitational/geometrical 
observables it is no longer possible~\cite{gacgrf98} 
to advocate an idealized description of the devices. 

Also the $T_{obs}$-dependent 
bound on the measurability of distances which I reviewed
in Section~3 encodes a
departure from ordinary Quantum Mechanics and 
a novel type of interplay
between system and measuring apparatus,
but the bound (\ref{hbound}) on the measurability of the amplitude of 
a gravity wave (which is one of the new results reported
in the present Article) should provide even stronger motivation
for the search of formalism in which Quantum Gravity 
is based on a new mechanics,
not exactly given by ordinary Quantum Mechanics.
In fact, while one might still hope
to find alternatives to the Salecker-Wigner measurement
procedure that allow to measure distances evading the 
bound (\ref{gacdl})
(or its $\delta D \sim max[S_d]$ version (\ref{sigmagactwothird})),
it appears hard to imagine that there could be 
anything (even among ``gedanken laboratories'')
better than an interferometer for measurements
of the amplitude of a gravity wave.

The fact that in the limit $\lambda^{o}_{gw} \rightarrow \infty$
(the no-gravity-wave limit) the bound (\ref{hbound}) reduces
to the bound $\delta h > \delta L /L$ is of course consistent
with the fact that when no gravity wave is going through the
interferometer the only Quantum-Gravity related noise sources
(if any) come directly from the distance fuzziness $\delta L$,
which I considered in the previous Sections.
The analysis reported in this Subsection appears to indicated
that the interferometer noise associated to distance fuzziness
could be simply seen as the $\lambda^{o}_{gw} \rightarrow \infty$
limit of a more complicated $\lambda^{o}_{gw}$-dependent
type of Quantum-Gravity related noise 
affecting the observation of gravity waves in a full Quantum Gravity
context.

It is also important to realize that the bound (\ref{hbound})
cannot be obtained by just assuming
that the Planck length $L_{planck}$
 provides the minimum uncertainty for lengths~\cite{garay}.
In fact, if the only limitation was $\delta D_L \ge L_{planck}$ the resulting
uncertainty on $h$, which I denote with $\delta h^{(L_{planck})}$,
would have the property
\begin{eqnarray}
min [\delta h^{(L_{planck})}] = min \left[ {L_{planck} \over 2 
\, \lambda^{o}_{gw}
\left| \sin \left( {L \over 2 \lambda^{o}_{gw}} 
\right) \right|} \right] = {L_{planck} \over 2 
\, \lambda^{o}_{gw}}
~,
\label{hboundlp}
\end{eqnarray}
whereas, exploiting the above-mentioned 
properties of the function $\sqrt{x}/|\sin (x/2)|$,
from (\ref{hbound}) one finds\footnote{I am here
(for ``pedagogical'' purposes) somewhat simplifying the comparison
between $\delta h$ and $\delta h^{(L_{planck})}$.
As mentioned, in principle one should take into account
both uncertainties inherent in the ``system'' under observation,
which are likely to be characterized exclusively by
the Planck-length bound, and uncertainties coming from 
the ``measuring apparatus'', which might easily involve
other length (or time) scales besides the Planck length.
It would therefore be proper to compare $\delta h^{(L_{planck})}$,
which would be the only contribution present in the conventional
idealization of ``classical devices'', with the 
sum $\delta h + \delta h^{(L_{planck})}$,
which, as appropriate for Quantum Gravity,
provides a sum of system-inherent uncertainties plus
apparatus-induced uncertainties.}
\begin{eqnarray}
min [\delta h]  > min \left[
{L_{planck} \over 2 \, \lambda^{o}_{gw}} 
{\sqrt{{ L/ \lambda^{o}_{gw} }} \over
\left| \sin \left( {L \over 2 \lambda^{o}_{gw}} 
\right) \right| } \right]
> min [\delta h^{(L_{planck})}] 
~.
\label{minhbound}
\end{eqnarray}
In general,  the dependence of $\delta h^{(L_{planck})}$
on $\lambda^{o}_{gw}$
is different from the one of $\delta h$.
Actually, in light of the comparison of
(\ref{hboundlp}) with (\ref{minhbound})
it is amusing to observe that
the bound (\ref{hbound}) could be seen as
the result of a minimum length $L_{planck}$ combined
with an $\lambda^{o}_{gw}$-dependent correction.
This would be consistent with some of the ideas mentioned
in Section~3, the energy-dependent
effect of {\it in vacuo} dispersion and the corresponding proposal
(\ref{deltagacgenfinalphaonesigmad}) for distance fuzziness,
in which the magnitude of the Quantum Gravity effect
depends rather sensitively on some energy-related 
aspect of the problem under investigation
(just like $\lambda^{o}_{gw}$ gives the energy of the
gravity wave).

It is easy to verify that
the bound (\ref{hbound}),
would not observably affect the operation of even the most 
sophisticated planned interferometers.
However, in the spirit of what I did in the previous Sections
considering the operative definition of distances, also
for the amplitudes of gravity waves the fact that we have encountered
an obstruction in the measurement analysis based on ordinary
Quantum Mechanics (and the fact that by mixing Gravity and Quantum
Mechanics we have obtained some intuition for novel qualitative
features of such gravity-wave amplitudes in Quantum Gravity)
could be used as starting point for
the proposal of novel Quantum Gravity effects
possibly larger than the estimate (\ref{hbound})
obtained by naive combination of Gravity and Quantum
Mechanics without any attempt at a fully Quantum-Gravity picture 
of the phenomenon.
Although possibly very interesting,
these fully Quantum-Gravity scenarios 
for the properties of gravity-wave amplitudes
will not be explored in the present Article.

\section{RELATIONS WITH OTHER QUANTUM $~~~~~~~~~~~~$
GRAVITY APPROACHES}

The general strategy for the search of Quantum Gravity
which has led to the arguments reviewed and/or presented in
the previous sections is evidently quite different from the strategy
adopted in other approaches to the unification of Gravity
and Quantum Mechanics. [I shall discuss these differences in greater
detail in Section~8.] However, it is becoming
increasingly clear (especially in discussions and research papers
that were motivated by Refs.~\cite{grbgac,gacgwi})
that in spite of these differences some common 
elements of intuition concerning the interplay of Gravity
and Quantum Mechanics are emerging.
In this Section I want to emphasize these relationships
with some Quantum Gravity approaches
and at the same time I want to clarify the differences
with respect to other Quantum Gravity approaches.

\subsection{Canonical Quantum Gravity}

One of the most popular Quantum Gravity approaches
(whose popularity might have been the reason
for the diffusion of the possibly
misleading name ``Quantum Gravity'')
is the one in which the ordinary canonical formalism
of Quantum Mechanics is applied to (some formulation of)
Einstein's Gravity.

While I must emphasize again~\cite{gacgrf98} that some of the
observations reviewed and/or reported in the previous sections
strongly suggest that Quantum Gravity should require a new mechanics,
not exactly given by ordinary Quantum Mechanics,
it is nonetheless encouraging that some of the phenomena
considered in the previous sections
have also emerged in 
studies of Canonical Quantum Gravity.

The most direct connection was found in the study
reported in Ref.~\cite{gampul},
which was motivated by Ref.~\cite{grbgac}.
In fact, Ref.~\cite{gampul} shows that the popular
Canonical/Loop Quantum Gravity~\cite{canoloop}
admits the phenomenon of deformed dispersion
relations, with the deformation going linearly with
the Planck length.

Concerning the bounds on the measurability of distances
it is probably fair to say that the situation
in Canonical/Loop Quantum Gravity is not yet clear
because the present formulations
do not appear to lead to
a compelling candidate ``length operator.''
This author would like to interpret
the problems associated
with the length operator as an indication that 
perhaps something unexpected
might actually emerge in Canonical/Loop Quantum Gravity
as a length operator,
possibly something with properties fitting the intuition
emerging from the analyses in Subsections~3.2,
3.3, and 3.6.
Actually, the random-walk 
space-time fuzziness models discussed in Subsection~3.3
might have a (somewhat weak, but intriguing) 
connection with ``Quantum Mechanics applied to Gravity''
at least to the level seen by 
comparison with the scenario discussed 
in Ref.~\cite{fotinilee}, which was motivated by 
the intuition that is emerging from investigations of 
the Canonical/Loop Quantum Gravity.
The ``moves'' of Ref.~\cite{fotinilee} share many of the properties
of the ``random steps'' of my random-walk models. 
Unfortunately, in both approaches
one is still searching for a more complete description of the 
dynamics, and particularly for estimates of how frequently (in time)
a $L_{planck}$-size step/move is taken.

\subsection{Non-commutative geometry and deformed symmetries}

Although this was not emphasized in the present Article,
some of the Quantum Gravity intuition emerging from the
observations in the previous sections fits rather naturally
within certain approaches based on non-commutative geometry and 
deformed symmetries.
In particular, there is growing evidence~\cite{gacxt,lukipap}
that theories living in
the non-commutative Minkowski space proposed
in Refs.~\cite{shahnkappamin,kpoin},
which involves a dimensionful (possibly Planck length related)
deformation parameter,
would host both the phenomenon of Planck-length-linear
deformations of dispersion relations and phenomena
setting ${T_{obs}}$-dependent bounds on the measurability of distances. 

In general, the possibility of 
dimensionful deformations of 
symmetries~\cite{kpoin,firenze} 
might be quite natural~\cite{gacgrf98}
if indeed the relation between system 
and measuring apparatus is modified at the Quantum Gravity level.
For example, 
the symmetries we observe 
in ordinary Quantum Mechanics experiments
at low energies might be the ones valid in the
limit in which the interaction between system and measuring
apparatus can be neglected.
The dimensionful parameter characterizing the deformation
of symmetries could mark a clear separation between
(high-energy) processes
in which the violations of ordinary symmetries are large
and (low-energy) processes in which ordinary symmetries 
hold to a very good approximation.

On the subject of quantum deformations
of space-time symmetries
interesting work has also been devoted 
(see, {\it e.g.}, Refs.~\cite{dopplicher,kempfmangano})
to frameworks that would
host a bound on the measurability of distances
of type (\ref{minlen}).

\subsection{Critical and non-critical String Theories}

Unfortunately, in the popular Quantum Gravity approach
based on Critical Superstring Theory\footnote{As already mentioned
the mechanics of String Theory
is just an ordinary Quantum Mechanics, the novelty of the approach
comes from the fact that the fundamental dynamical entities
are  extended objects rather than point particles.} 
not many results have been derived 
concerning directly the quantum properties of space-time.
Perhaps the most noticeable such results are the ones on
limitations on the measurability of distances emerged in the
scattering analyses reported in Refs.~\cite{venekonmen,dbrscatt},
which I already mentioned in Subsection~3.1, since they
provide support for the hypothesis that also
Critical Superstring Theory
might host a bound on the measurability of distances
of type (\ref{minlen}).

A rather different picture is emerging (within the difficulties
of this rich formalism) in Liouville (non-critical) String 
Theory~\cite{emn}, whose development was partly motivated 
by intuition concerning the ``Quantum Gravity vacuum''
that is rather close to the one traditionally associated
to the works of Wheeler~\cite{wheely} and Hawking~\cite{hawk}.
Evidence has been found~\cite{aemn1} 
in Liouville String Theory
supporting the validity  
of deformed dispersion relations, with the 
deformation going linearly with the Planck/string length.
In the sense clarified in Subsections~3.4 this approach might
also host a bound on the measurability of distances which grows
with $\sqrt{T_{obs}}$.

\subsection{Other types of measurement analyses}

In light of the scarce opportunities to get
any experimental input in the search for Quantum Gravity,
it is not surprising that
many authors have been seeking some
intuition by formal analyses of
the ways in which
the interplay between Gravity and Quantum Mechanics
could affect measurement procedures.
A large portion of these analyses
produced a ``$min [\delta D]$'' with $D$ denoting
a distance; however, the same type of notation
was used for structures defined in significantly
different manner.
Also different meanings have been given by different
authors to the statement ``absolute bound on the measurability
of an observable.''
Quite important for the topics here discussed
are the differences (which might not be
totally transparent as a result of this unfortunate
choice of overlapping notations)
between the approach advocated 
in the present Article 
(and in Refs.~\cite{gacgwi,gacmpla,gacgrf98}) 
and the approaches advocated in Refs.~\cite{wign,diosi,ng,karo}.
In the present Article ``$min [\delta D]$'' denotes
an absolute limitation on the measurability of a distance $D$.
The studies~\cite{wign,diosi,karo} 
analyzed the interplay of Gravity and Quantum Mechanics
in defining a net of time-like geodesics, and 
in those studies ``$min [\delta D]$''
characterizes the maximum ``tightness'' achievable
for the net of time-like geodesics.
Moreover, 
in Refs.~\cite{wign,diosi,ng,karo} it was required that
the measurement procedure should not affect/modify the
geometric observable being measured, 
and ``absolute bounds on the measurability''
were obtained in this specific sense.
Instead, here and in Refs.~\cite{gacmpla,gacgrf98} I allowed the
possibility for the observable which is being measured to depend
also on the devices (the underlying view is that observables
in Quantum Gravity would always be, in a sense, shared properties
of ``system'' and ``apparatus''),
and I only required that
the nature of the devices be consistent with the various
stages of the measurement procedure ({\it e.g.}, a black-hole
device would not allow some of the required exchanges of signal).
My measurability bounds are therefore to be intended from this
more fundamental perspective, and this is crucial for the
possibility that these measurability bounds be associated
to a fundamental Quantum-Gravity mechanism for ``fuzziness''
(quantum fluctuations of space-time).
The analyses reported
in Refs.~\cite{wign,diosi,ng,karo} 
did not include any reference to fuzzy space-times
of the type operatively defined in Section~2.

The more fundamental nature of the bounds I obtained
is also crucial for the arguments suggesting~\cite{gacmpla,gacgrf98}
that Quantum Gravity might require a new mechanics,
not exactly given by ordinary Quantum Mechanics.
The analyses reported
in Refs.~\cite{wign,diosi,ng,karo} 
did not include any reference to this possibility.

I also notice that the conjectured relation between 
measurability bounds and 
noise levels in interferometers ({\it e.g.} the ones
characterized by $S(f) \sim f^{-1}$ or $S(f) \sim f^{-5/6}$)
is based on the dependence of the measurability bounds on the time of
observation $T_{obs}$. In fact, this  $T_{obs}$-dependence
has been here emphasized,
while in Refs.~\cite{diosi,ng,karo} 
the emphasis was placed on observed lengths rather than on the
time needed to observe them.

Having clarified that there is 
a ``double difference'' (different ``$min$'' and 
different ``$\delta D$'')
between 
the meaning of $min [\delta D]$ 
adopted in the present Article 
and the meaning of $min [\delta D]$ adopted
in Refs.~\cite{wign,diosi,ng,karo}, it is however important to 
notice that the studies reported
in Refs.~\cite{diosi,ng,karo} 
were among the first studies which showed how in some aspects
of measurement analysis the Planck length might appear
together with other length scales in the problem.
For example, a Quantum Gravity effect 
naturally involving something
of length-squared dimensions might not necessarily 
go like $L_{planck}^2$, in some case it could go
like $\Lambda L_{planck}$, with $\Lambda$ some other 
length scale in the problem.
Some of my arguments are examples of this possibility;
in particular, I find in some cases relations of the type
(see, {\it e.g.}, Eq.~(\ref{deltawignOLD}))
\begin{eqnarray}
\delta D \geq 
\delta x^* 
+ { A \over \delta x^* } \geq \sqrt{A}
~,
\label{example}
\end{eqnarray}
where $A$, which has length-squared dimensions,
turns out to be given by the product of
the $L_{planck}$-like small fundamental length $L_{QG}$
and the typically larger length scale $c T_{obs}$.

Interestingly, the
analysis of the interplay of Gravity and Quantum Mechanics
in defining a net of time-like geodesics reported in Ref.~\cite{diosi}
concluded that the maximum ``tightness'' achievable 
for the geodesics would be characterized
by $\sqrt{L_{planck}^2 R^{-1} s}$,
where $R$ is the radius of the (spherically symmetric) clocks
whose world lines define the network of geodesics,
and $s$ is the characteristic distance scale 
over which one is intending to define such a network.
The $\sqrt{L_{planck}^2 R^{-1} s}$ maximum tightness discussed 
in Ref.~\cite{diosi} is formally analogous to my Eq.~(\ref{deltagacdlbis}),
but, as clarified above, this ``maximum tightness''
was defined in a way that is very (``doubly'')
different from my ``$min [\delta D]$'',
and therefore the two proposals have 
completely different physical implications.
Actually, in Ref.~\cite{diosi} it was also stated that 
for a single geodesic distance (which
might be closer to the type of distance 
measurability analysis reported here and in
Refs.~\cite{gacmpla,gacgrf98}) one could achieve accuracy
significantly better than the formula $\sqrt{L_{planck}^2 R^{-1} s}$,
which was interpreted in Ref.~\cite{diosi} 
as a direct result of the structure of a network of geodesics.

Relations of the type $min [\delta D] \sim (L_{planck}^2 D)^{(1/3)}$,
which are formally analogous to Eq.~(\ref{deltagactwothird}),
were encountered in the analysis of 
maximum tightness achievable for 
a geodesics network reported in Ref.~\cite{karo} 
and in the analysis of measurability of distances 
reported in Ref.~\cite{ng}.
Although once again the definitions of ``$min$''
and ``$\delta D$'' used in these studies
are completely different from the ones relevant for 
the ``$min [\delta D]$'' of Eq.~(\ref{deltagactwothird}),
the analyses reported in Ref.~\cite{ng,karo} do provide
some additional motivation for the scenario (\ref{deltagactwothird}),
at least in as much as they give
examples of the fact that behaviour of 
the type $L_{planck}^{2/3}$ can naturally emerge in 
Quantum-Gravity measurement analyses.

\subsection{Other interferometry-based Quantum-Gravity studies}

Several authors have put forward ideas which combine in one
or another way some aspects of interferometry 
and candidate Quantum Gravity phenomena.
While the viewpoints and the results of all of these works
are significantly different from the ones of the present Article,
it seems appropriate to at least mention briefly these studies,
for the benefit of the interested reader.

A first example, on which I shall return in the next Section,
is provided by the idea~\cite{venegwi}
that we might be able to use modern gravity-wave interferometers
to investigate certain
candidate early-universe String Theory effects.

The studies reported in Ref.~\cite{peri} (and references
therein) have considered how certain effectively
stochastic properties of space-time would affect
the evolution of quantum-mechanical states.
The stochastic properties there considered are different from
the ones discussed in the present Article, but were introduced
within a similar viewpoint, {\it i.e.} stochastic processes
as effective description of quantum space-time processes.
The implications of these stochastic properties
for the evolution of quantum-mechanical states
were modeled via the formalism of ``primary  state diffusion'',
but only rather crude models turned out to be treatable.
Atom interferometers were found to have properties suitable
for tests of this scenario. I should however emphasize that
in Ref.~\cite{peri} the proposed tests concerned the
Quantum Mechanics of systems leaving in a fuzzy space-time,
whereas here and in Ref.~\cite{gacgwi} I have discussed
direct tests of effectively stochastic properties of
space-time.

The studies reported in Refs.~\cite{jare1,jare2}
are more closely related to the physics of gravity-wave
interferometers. In particular, combining a detailed analysis
of certain aspects of interferometry and the assumption that
quantum space-time effects could be estimated using
ordinary Quantum Mechanics applied to Einstein's gravity,
Refs.~\cite{jare1,jare2} developed a model of
Quantum-Gravity induced noise for interferometers
which fits within the scenario I here discussed
in Subsection~3.1. [Actually, Refs.~\cite{jare1,jare2} discuss
in greater detail the spectral features encoded in
Eqs.~(\ref{no1})-(\ref{no1spectrum}), while, as explained in
Subsection~3.1, it was for me sufficient to provide a simplified
discussion.]
As mentioned in Subsection~3.1, it is not surprising that
the assumption that Quantum Gravity be given by an
ordinary Quantum Mechanics applied to
(some formulation of) Einstein's gravity
would lead to noise levels of the type
encoded in Eqs.~(\ref{no1})-(\ref{no1spectrum}).

The recent paper Ref.~\cite{camacho} proposed
certain quantum properties of gravity waves
and discussed the implications for gravity-wave
interferometry. 
Let me emphasize that instead
the effects considered here and in Ref.~\cite{gacgwi}
concern the properties of the interferometer and 
would affect the operation of any interferometer
whether or not it would be used to detect gravity waves.
Here and in Ref.~\cite{gacgwi}
the emphasis on modern gravity-wave interferometers
is only due to the fact that these interferometers,
because of the extraordinary challenges posed by
the detection of classical gravity waves, are the most advanced
interferometers available and therefore provide the
best opportunity to test scenarios for
Quantum-Gravity induced noise in interferometers.

\section{A QUANTUM-GRAVITY PHENOMENOLOGY $~~~~~$
PROGRAMME}

While opportunities to test experimentally
the nature of the interplay between
Gravity and Quantum Mechanics
remain extremely rare,
the proposals now 
available~\cite{elmn,venegwi,grbgac,ahluexp,cow,gacgwi}
represent a small fortune with respect to the
expectations of not many years ago.
We have finally at least reached the point that
the most optimistic/speculative estimates of Quantum Gravity
effects can be falsified.
In searching for even more opportunities to
test Quantum Gravity it is useful to analyze the proposals put forward
in Refs.~\cite{elmn,venegwi,grbgac,ahluexp,cow,gacgwi}
as representatives of the two generic mechanisms
that one might imagine to use in Quantum-Gravity experiments.
Let me comment here on these mechanisms.
The most natural discovery strategy would of course resort to 
strong Quantum Gravity effects, of the type we expect for
collisions of elementary particles endowed with momenta of order 
the Planck mass ($10^{19}GeV$).
Since presently and for the foreseeable future
we do not expect to be able to set up
such collisions, 
the only opportunities to find evidence of strong Quantum Gravity effects
should be found in natural phenomena 
({\it e.g.} astrophysical contexts
that might excite strong Quantum Gravity effects)
rather than in controlled laboratory setups.
An example is provided by the experiment proposed 
in Ref.~\cite{venegwi} which would be looking for residual
traces of some strong Quantum Gravity effects\footnote{Most
of the effects considered in Ref.~\cite{venegwi}
actually concern the interplay between classical Gravity
and Quantum Mechanics, so they pertain to a very special
regime of Quantum Gravity. This is also true of the
experiments on gravitationally induced quantum
phases~\cite{ahluexp,cow,chu}.
Instead the experiments discussed here and in
Refs.~\cite{elmn,grbgac,gacgwi} concern
proposed quantum properties of space-time itself,
and could therefore probe even more deeply the
structure of Quantum Gravity.}
(specifically, Critical Superstring Theory
effects\cite{stringcosm})
which might have occurred in the early Universe.

Another class of Quantum Gravity experiments 
is based on physical contexts in which small Quantum Gravity effects
lead to observably large signatures thanks to the
interplay with a naturally large number present in such contexts.
This is the basic mechanism underlying all the proposals
in Refs.~\cite{elmn,grbgac,ahluexp,cow}
and underlying the interferometric studies of space-time fuzziness
proposed in Ref.~\cite{gacgwi} 
which I have here discussed in detail.
For the interferometric studies which I am proposing
the large number is essentially provided by 
the ratio between the inverse of the Planck time and the 
typical frequencies of operation of gravity-wave interferometers.
In practice if some of the space-time fuzziness scenarios
discussed in Section~3 capture actual features of quantum space-time,
in a time as long as the inverse of the 
typical gravity-wave interferometer frequency of operation
an extremely large number of minute quantum fluctuations in the
distance $D$ could add up. A large sum of small quantities can give 
a sizeable final result, and in fact this final result
would be observable if for example the noise induced by fuzziness was 
characterised by $f^{-2} c L_{QG}$ which is comparable in
size to the corresponding quantity $[S_{exp}(f)]^2$ characterising the 
noise levels achievable with modern interferometers.

For the physical context of gamma rays reaching us from far away
astrophysical objects the large number can be provided by the
ratio between the time travelled by the gamma rays
and the time scale over which the signal presents significant structure
(time spread of peaks etc.). 
The proposal made in Ref.~\cite{grbgac}
basically uses the fact that this allows to add up a very large
number of very
minute dispersion-inducing Quantum Gravity effects, and if the
deformation of the dispersion relation
goes linearly with the Planck length the resulting 
energy-dependent time-delay
turns out to be comparable to the time scale that characterizes some
of these astrophsical signals, thereby allowing a direct test
of the Quantum Gravity scenario.

Similarly, experiments investigating the quantum phases 
induced by large gravitational fields~\cite{ahluexp,cow,chu}
(the only aspect of the interplay between Gravity and Quantum Mechanics 
on which we already have positive ``discovery'' data~\cite{cow,chu}) 
exploit the fact that gravitational forces are additive
and therefore, for example, gravitational effects due to the earth
are the result of a very large number of very
minute gravitational effects (instead we would not be 
able to measure the quantum phases 
induced by a single elementary particle).

The large number involved in the
possibility that Quantum Gravity effects might leave an
observable trace
in some aspects of the phenomenology of
the neutral-kaon system~\cite{elmn} 
cannot be directly interpreted 
as a the number of minute Quantum Gravity effects
to which the system is exposed.
It is rather that the conjectured Quantum Gravity effects
would involve in addition to the small dimensionless
ratio between the energy of the
kaons and the Planck energy also a very large
dimensionless ratio~\cite{elmn} 
characterising the physics of neutral kaons.

This idea of figuring out ways to put together many minute effects
(which until a short time ago had been strangely dismissed by
the Quantum Gravity community)
has a time honored tradition in physics. 
Perhaps the clearest example
is the particle-physics experiment setting bounds on proton
lifetime. The relevant dimensionless ratio 
characterising proton-decay analyses
is extremely small
(somewhere in the neighborhood of $10^{-64}$,
since it is given by
the fourth power of the ratio between the mass of the proton
and the grandunification scale),
but by keeping under observation
a correspondingly large number of protons experimentalists are
managing\footnote{This author's familiarity~\cite{gactesi}
with the accomplishments of proton-decay experiments has
certainly contributed to the moderate optimism for the outlook
of Quantum Gravity phenomenology which is implicit in the
present Article.}
to set highly significant bounds.

Another point of contact between proposed Quantum Gravity experiments
and proton decay experiments is that a crucial role in rendering
the experiment viable is the fact that the process under investigation
would violate some of the symmetries of ordinary physics.
This plays a central role in the experiments proposed in
Refs.~\cite{elmn,grbgac}.

\section{MORE ON A LOW-ENERGY EFFECTIVE $~~~~~~~~~$
THEORY OF QUANTUM GRAVITY}

While the primary emphasis has been on direct experimental tests of
crude scenarios for space-time fuzziness, part of this Article
has been devoted to the discussion (expanding on what was
reported in Refs.~\cite{gacmpla,gacgrf98})
of the properties that one could demand of a theory
suitable for a first stage of partial
unification of Gravity and Quantum Mechanics.
This first stage of partial unification
would be a low-energy effective theory capturing
only some rough features of Quantum Gravity,
possibly associated with
the structure of the non-trivial ``Quantum Gravity vacuum''.

One of the features that appear desirable for
an effective low-energy theory
of Quantum Gravity is that its mechanics be
not exactly given by ordinary Quantum Mechanics.
I have reviewed some of the arguments~\cite{gacmpla,gacgrf98}
in support of this hypothesis when I discussed the
Salecker-Wigner setup for the measurement of distances,
and showed that the problems associated with
the infinite-mass classical-device limit provide encouragement
for the idea that the analysis of
Quantum Gravity experiments 
should be fundamentally different from the one
of the experiments described by ordinary Quantum Mechanics.
A similar conclusion was already drawn in the context
of attempts (see, {\it e.g.}, Ref.~\cite{bergstac})
to generalize to
the study of the measurability of gravitational fields
the famous Bohr-Rosenfeld analysis~\cite{rose}
of the measurability of the electromagnetic field.
In fact, in order to achieve the accuracy allowed by
the formalism of ordinary Quantum Mechanics,
the Bohr-Rosenfeld measurement procedure resorts to
ideal test particles of infinite mass, which would of course
not be admissible probes in a gravitational
context~\cite{bergstac}. 
Since all of
the (extensive) experimental evidence for ordinary
Quantum Mechanics
comes from experiments in which the behaviour of the devices
can be meaningfully approximated as classical,
and moreover it is well-understood
that the conceptual structure of ordinary
Quantum Mechanics makes it only acceptable as the theoretical
framework for the description of the outcomes of this
specific type of experiments,
it seems reasonable to explore the possibility that
Quantum Gravity might require a new mechanics,
not exactly given by ordinary Quantum Mechanics
and probably involving a novel
(in a sense, ``more democratic'')
relationship between ``measuring apparatus''
and ``system''. 

Other (related) plausible features
of the correct effective low-energy theory
of Quantum Gravity are
novel bounds on the measurability of distances.
This appears to be an inevitable consequence of
relinquishing the idealized methods of
measurement analysis that rely
on the artifacts of the infinite-mass
classical-device limit.
If indeed one of these novel measurability
bounds holds in the physical world, and if indeed the
structure of the Quantum-Gravity vacuum
is non-trivial and involves space-time fuzziness,
it appears also plausible that this two features be related,
{\it i.e.} that the fuzziness of space-time would be
ultimately responsible for the measurability bounds.
It is this scenario which I have investigated here
and in Ref.~\cite{gacgwi}, emphasizing the
opportunity for direct tests which is provided by modern
interferometers.

The intuition emerging from these first investigations
of the properties of a low-energy effective Quantum Gravity
might or might not turn out to be accurate,
but additional work on this first stage of partial unification
of Gravity and Quantum Mechanics is anyway well motivated
in light of the huge gap between the Planck regime
and the physical regimes
ordinarily accessed in present-day particle-physics or
gravity experiments.
Results on a low-energy effective Quantum Gravity might
provide a  perspective on Quantum Gravity
that is complementary with respect to the
one emerging from approaches based on proposals
for a one-step full unification of Gravity and
Quantum Mechanics.
On one side of this complementarity there are the
attempts to find a low-energy effective Quantum Gravity
which are necessarily driven by intuition based on
direct extrapolation from known physical regimes;
they are therefore rather close to the phenomelogical realm
but they are confronted by huge difficulties when
trying to incorporate the physical intuition within a
completely new formalism.
On the other side there are the attempts of
one-step full unification of Gravity and Quantum Mechanics,
which usually start from some intuition concerning the
appropriate formalism ({\it e.g.}, ``Canonical/Loop
Quantum Gravity''~\cite{canoloop} or ``Critical
Superstring Theory''~\cite{dbrane,critstring})
but are confronted by huge difficulties when
trying to ``come down'' to the level of phenomenological
predictions.
These complementary perspectives might
meet at the mid-way point leading
to new insight in Quantum Gravity physics.
One instance in which this mid-way-point meeting 
has already been successful is provided by the
candidate phenomenon of Quantum-Gravity induced
deformed dispersion relations,
which was proposed within a purely phenomenological
analysis~\cite{grbgac} of the type needed for
the search of a low-energy theory of Quantum Gravity,
but was then shown~\cite{gampul} to be consistent with
the structure of Canonical/Loop Quantum Gravity.

\section{OUTLOOK}

The panorama of opportunities for
Quantum Gravity phenomenology
is certainly becoming richer.
In this Article I
have taken the conservative viewpoint
that the length scales parametrizing
proposed Quantum Gravity phenomena
should be somewhere in the neighborhood
of the Planck length, but I have taken
the optimistic (although supported
by various Quantum Gravity scenarios,
including Canonical/Loop
Quantum Gravity~\cite{canoloop,gampul})
viewpoint that there should be Quantum
Gravity effects going linearly or
quadratically with the Planck length,
{\it i.e.} effects which are penalized only by one
or two powers of the Planck length.

An exciting recent development is that results
in the general area of String Theory have motivated
work (see, {\it e.g.}, Ref.~\cite{largextra})
on theories with large extra dimensions
in which rather naturally Quantum Gravity
effects would become significant at scales much
larger than the conventional Planck length.
In such scenarios one expects to find 
phenomena for which the length scale characterizing the
onset of large Quantum-Gravity corrections is much
larger than the conventional Planck length.

The example of advanced modern
interferometers here emphasized
provides further evidence (in addition to the one emerging
from Refs.~\cite{elmn,grbgac}) of the fact that
we should eventually be able to find signatures
of Quantum Gravity effects if they
are linear in the conventional Planck length.
If the physical world only hosts effects that are quadratic
in the deformation length scale,
values of this length scale of order the Planck length
would probably be out of reach for the foreseeable future,
but effects quadratic in the larger length scales
characterizing scenarios of the type in Ref.~\cite{largextra}
might be experimentally accessible.

On the theory side an exciting opportunity for future research
appears to be provided by the possibility of
exchanges of ideas between
the more phenomenological/intuitive studies 
appropriate for the search of a low-energy effective Quantum Gravity
and the more rigorous/formal studies used in searches of
fully consistent Quantum Gravity theories.
As mentioned at the end of the preceding Section,
the first example of such an exchange has led to the exciting
realization that deformed dispersion relations
linear in the Plack length appear plausible~\cite{grbgac,gacgrf98,aemn1}
both
from the point of view of heuristic 
phenomenological analyses and are also a rather general prediction
of Canonical/Loop Quantum Gravity~\cite{gampul}.
Additional exchanges of this type appear likely.
For example, the intuition coming from the 
low-energy effective Quantum Gravity viewpoint
on distance fuzziness which I discussed here
might prove useful for those Quantum Gravity approaches 
(again an example is provided by Canonical/Loop Quantum Gravity)
in which there is substantial evidence of space-time
fuzziness but one has not yet achieved a satisfactory
description of fuzzy distances.

\vglue 0.6cm
\leftline{\Large {\bf Acknowledgements}}
\vglue 0.4cm
I owe special thanks to Abhay Ashtekar, since he 
suggested to me that gravity-wave interferometers might be 
useful for experimental tests of some of the Quantum-Gravity 
phenomena that I have been investigating.
My understanding of Refs.~\cite{elmn} and \cite{venegwi} 
benefited from conversations 
with N.E.~Mavromatos and G.~Veneziano.
I am also happy to acknowledge
a kind email message from A.~Camacho which provided
positive feed-back on my Ref.~\cite{gacgwi} and also
made me aware of the works in Refs.~\cite{jare1,jare2,camacho}.
Still on the ``theory side'' I am grateful to several colleagues 
who provided encouragement and stimulating feed-back, 
particularly D.~Ahluwalia, J.~Ellis, J.~Lukierski,
C.~Rovelli, S.~Sarkar, L.~Smolin and J.~Stachel.
On the ``experiment side'' I would like to thank
F.~Barone, 
J.~Faist, 
R.~Flaminio,
L.~Gammaitoni, 
T.~Huffman,
L.~Marrucci
and
M.~Punturo
for useful conversations 
on various aspects of interferometry.

\newpage
\baselineskip 12pt plus .5pt minus .5pt

\end{document}